\def\aa{{A\&A}}

\def\apj{{ApJ}}  
\def\apjs{{ApJS}}  
  
\def\mnras{{MNRAS}}  
\def\nat{{Nature}}

\newcommand{\be}{\begin{equation}}  
\newcommand{\ee}{\end{equation}}  
\newcommand{\Deln}{\ensuremath{\Delta N_\nu\;}}  
\def\Nnu{$N_{\nu}$~}  
\newcommand{\epm}{\ensuremath{e^{\pm}\;}}  
  
\def\ie{{i.e.},~}  
\def\eg{{e.g.},~}  
\def\etal{{et al.}~}  
\def\4he{$^4$He}  
\def\3he{$^3$He}  
\def\7li{$^7$Li}  
\def\Yp{Y$_{\rm P}$~}  
\def\yd{$y_{\rm D}$~}  
\def\y3{$y_{3}$~}  
\def\hii{H\thinspace{$\scriptstyle{\rm II}$}~}  
\def\hi{H\thinspace{$\scriptstyle{\rm I}$}~}  
\def\di{D\thinspace{$\scriptstyle{\rm I}$}~}  
  
\newcommand\la{\lower0.6ex\vbox{\hbox{\ensuremath{\buildrel{\textstyle<}\over{\sim}\  
}}}}  
\newcommand\ga{\lower0.6ex\vbox{\hbox{\ensuremath{\buildrel{\textstyle>}\over{\sim}\  
}}}}

\newcommand{\omb}{\ensuremath{\Omega_{\rm b}\;}}

\documentclass[cup5b]{caps}  
\usepackage{graphicx}  
\usepackage{amssymb}  
\usepackage{ociwsymp2}   
  
\HeadText{G. Steigman}

\begin{document}  
  
\pagenumbering{arabic}  
  
\author[]{GARY STEIGMAN\\Departments of Physics and of Astronomy,   
The Ohio State University}  
  
\chapter{Big Bang Nucleosynthesis: \\ Probing the First 20 Minutes}  
  
\begin{abstract}  
Within the first 20 minutes of the evolution of the hot, dense, early  
Universe, astrophysically interesting abundances of deuterium, helium-3,   
helium-4, and lithium-7 were synthesized by the cosmic nuclear reactor.  
The primordial abundances of these light nuclides produced during Big Bang  
Nucleosynthesis (BBN) are sensitive to the universal density of baryons   
and to the early-Universe expansion rate which at early epochs is   
governed by the energy density in relativistic particles (``radiation'')   
such as photons and neutrinos.  Some 380 kyr later, when the cosmic   
background radiation (CBR) radiation was freed from the embrace of the   
ionized plasma of protons and electrons, the spectrum of temperature   
fluctuations imprinted on the CBR also depended on the baryon and   
radiation densities.  The comparison between the constraints imposed   
by BBN and those from the CBR reveals a remarkably consistent picture   
of the Universe at two widely separated epochs in its evolution.    
Combining these two probes leads to new and tighter constraints on the   
baryon density at present, on possible new physics beyond the standard   
model of particle physics, as well as identifying some challenges to  
astronomy and astrophysics.  In this review the current status of BBN  
will be presented along with the associated estimates of the baryon  
density and of the energy density in radiation.   
\end{abstract}  
  
\section{Introduction}  
  
The present Universe is observed to be expanding and filled with radiation   
(the 2.7~K cosmic background radiation; CBR) as well as with ``ordinary   
matter'' (baryons), ``dark matter,'' and ``dark energy.''  As a consequence,   
the early Universe must have been hot and dense. Sufficiently early in   
its evolution, the universal energy density would have been dominated by   
relativistic particles (``radiation dominated'').  During its early evolution   
the Universe passed through a brief epoch when it functioned as a cosmic   
nuclear reactor, synthesizing the lightest nuclides: D, \3he, \4he, and \7li.   
These relics from the distant past provide a unique window on the  early  
evolution of the Universe, as well as being valuable probes of the standard   
models of cosmology and particle physics. Comparing the predicted primordial   
abundances with those inferred from observational data tests the standard   
models and may uncover clues to their modifications and/or to extensions   
beyond them.  It is clear that Big Bang Nucleosynthesis (BBN), one of the   
pillars of modern cosmology, has a crucial role to play as the study of   
the evolution of the Universe enters a new, data-rich era.  
  
As with all science, cosmology depends on the interplay between theoretical   
ideas and observational data.  As new and better data become available,  
models may need to be refined, revised, or even replaced.  A consequence  
of this is that any {\it review} such as this one is merely a signpost   
along the road to a better understanding of our Universe.  While details  
of the current ``standard'' model, along with some of its more popular  
variants to be discussed here, may need to be revised or rejected in the  
future, the underlying physics to be described here can provide a useful  
framework and context for understanding those changes.  Any quantitative   
conclusions to be reached today will surely need to be modified in the   
light of new data.  This review is, then, a status report on the standard   
model, highlighting its successes as well as exposing the current challenges   
it faces.  While we may rejoice in the consistency of the standard model,   
there is still much work, theoretical and observational, to be done.  
  
\section{An Overview of BBN}  
  
To set a context for the confrontation of theoretical predictions with  
observational data it is useful to review the physics and cosmology of  
the early evolution of the Universe, touching on the specifics relevant  
for the synthesis of the light nuclides during the first $\sim 20$~minutes.  
In this section is presented an overview of this evolution along with   
the predicted primordial abundances, first in the standard model and   
then for two examples of nonstandard models which involve variations   
on the early-Universe expansion rate (Steigman, Schramm, \& Gunn 1977)
or asymmetries between the number of neutrinos and antineutrinos  
(\eg Kang \& Steigman 1992, and references therein).  
  
\subsection{Early Evolution}  
  
Discussion of BBN can begin when the Universe is a few tenths of a second   
old and the temperature is a few MeV.  At such an early epoch the energy   
density is dominated by the relativistic (R) particles present, and the   
Universe is said to be ``radiation-dominated.''  For sufficiently early 
times, when the temperature is a few times higher than the electron 
rest-mass energy, these are photons, \epm pairs, and, for the standard 
model of particle physics, three flavors of left-handed (\ie one helicity 
state) neutrinos (and their right-handed antineutrinos).  
\be  
\rho_{\rm R} = \rho_{\gamma} + \rho_{e} +  
3\rho_{\nu} = {43 \over 8}\rho_{\gamma},  
\label{eq:rho0}  
\ee  
where $\rho_{\gamma}$ is the energy density in CBR photons (which,  
today, have redshifted to become the CBR photons at a temperature  
of 2.7 K).    
  
In standard BBN (SBBN) it is assumed that the neutrinos are fully   
decoupled prior to \epm annihilation and do not share in the energy   
transferred from the annihilating \epm pairs to the CBR photons.    
In this approximation, in the post-\epm annihilation Universe, the   
photons are hotter than the neutrinos by a factor $T_{\gamma}/T_{\nu}   
= (11/4)^{1/3}$, and the relativistic energy density is  
\be  
\rho_{\rm R} = \rho_{\gamma} + 3\rho_{\nu} = 1.68\rho_{\gamma}.  
\ee  
  
During these radiation-dominated epochs the age ($t$) and the energy density 
are related  
by ${32\pi G \over 3}\rho_{\rm R} t^{2} = 1$, so that once the particle   
content ($\rho_{\rm R}$) is specified, the age of the Universe is known   
(as a function of the CBR temperature $T_{\gamma}$).  In the standard model,  
\be  
{\rm Pre-}\epm {\rm annihilation}:~~t~T_{\gamma}^{2} = 0.738~{\rm MeV^{2}~s},  
\label{eq:ttpre}  
\ee  
\be  
{\rm Post-}\epm {\rm annihilation}:~~t~T_{\gamma}^{2} = 1.32~{\rm MeV^{2}~s}.  
\label{eq:ttpost}  
\ee  
  
Also present at these early times are neutrons and protons,   
albeit in trace amounts compared to the relativistic particles.    
The relative abundance of neutrons and protons is determined   
by the charged-current weak interactions.   
\be   
p + e^{-} ~\longleftrightarrow ~n + \nu_{e}\,, ~~~~n + e^{+}   
~\longleftrightarrow ~p + \bar{\nu}_{e}\,, ~~~~n ~\longleftrightarrow    
~p + e^{-} + \bar{\nu}_{e}\,.   
\label{eq:betadecay}   
\ee   
As time goes by and the Universe expands and cools, the lighter   
protons are favored over the heavier neutrons and the neutron-to-proton   
ratio decreases, initially following the equilibrium form $(n/p)_{eq}   
\propto $~exp$(-\Delta m/T)$, where $\Delta m = 1.29$ MeV is the   
neutron-proton mass difference.  As the temperature drops the 
two-body collisions in Equation~\ref{eq:betadecay} become too slow 
to maintain equilibrium and the neutron-to-proton ratio, while 
continuing to decrease, begins to deviate from ({\it exceeds})   
this equilibrium value.  For later reference, we note that if 
there is an {\it asymmetry} between the numbers of $\nu_{e}$ 
and $\bar\nu_{e}$ (``neutrino degeneracy''), described by a 
chemical potential $\mu_{e}$ (such that for $\mu_{e} > 0$ there 
are more $\nu_{e}$ than $\bar\nu_{e}$), then the equilibrium 
neutron-to-proton ratio is modified to $(n/p) \propto~ \exp
(-\Delta m/T - \mu_{e}/T)$. In place of the neutrino chemical 
potential, it is convenient to introduce the dimensionless   
degeneracy parameter $\xi_{e} \equiv \mu_{e}/T$, which is 
invariant as the Universe expands.    
  
Prior to \epm annihilation, at $T \approx 0.8$~MeV when the Universe   
is $\sim 1$~second old, the two-body reactions regulating $n/p$   
become slow compared to the universal expansion rate and this ratio   
``freezes in,'' although, in reality, it continues to decrease, albeit   
more slowly than would be the case for equilibrium. Later, when the   
Universe is several hundred seconds old, a time comparable to the   
neutron lifetime ($\tau_{n} = 885.7 \pm 0.8$~s), the $n/p$ ratio   
resumes falling exponentially: $n/p \propto $~exp$(-t/\tau_{n})$.   
Since there are several billion CBR photons for every nucleon (baryon),   
the abundances of any complex nuclei are entirely negligible at these   
early times.  
  
Notice that since the $n/p$ ratio depends on the competition between the   
weak interaction rates and the early-Universe expansion rate (as well as   
on a possible neutrino asymmetry), any deviations from the standard model   
(\eg $\rho_{\rm R} \rightarrow \rho_{\rm R} + \rho_{X}$ or $\xi_{e} \neq 
0$) will change the relative numbers of neutrons and protons available 
for building more complex nuclides.    
  
\subsection{Building the Elements}   
   
At the same time that neutrons and protons are interconverting, they are also   
colliding among themselves to create deuterons: $n + p \longleftrightarrow   
D + \gamma$.  However, at early times, when the density and average energy   
of the CBR photons are very high, the newly formed deuterons find themselves   
bathed in a background of high-energy gamma rays capable of photodissociating   
them.  Since there are more than a billion photons for every nucleon in the   
Universe, before the deuteron can capture a neutron or a proton to begin   
building the heavier nuclides, the deuteron is photodissociated.  This   
bottleneck to BBN persists until the temperature drops sufficiently so that   
there are too few photons energetic enough to photodissociate the deuterons   
before they can capture nucleons to launch BBN.  This occurs after \epm   
annihilation, when the Universe is a few minutes old and the temperature   
has dropped below 80 keV (0.08 MeV).   
   
Once BBN begins in earnest, neutrons and protons quickly combine to form  
D, $^3$H, \3he, and \4he.  Here, at \4he, there is a different kind of  
bottleneck.  There are no stable mass-5 nuclides.  To jump this gap 
requires \4he reactions with D or $^3$H or \3he, all of which are positively  
charged.  The Coulomb repulsion among these colliding nuclei suppresses  
the reaction rates, ensuring that virtually all of the neutrons available  
for BBN are incorporated in \4he (the most tightly bound of the light  
nuclides), and also that the abundances of the heavier nuclides are  
severely depressed below that of \4he (and even of D and \3he).  Recall  
that $^3$H is unstable, decaying to \3he.  The few reactions that manage  
to bridge the mass-5 gap lead mainly to mass-7 (\7li or $^7$Be, which,
later, when the Universe has cooled further, will capture an electron  
and decay to \7li); the abundance of $^6$Li is 1 to 2 orders of  
magnitude below that of the more tightly bound \7li.  Finally, there  
is another gap at mass-8.  This absence of any stable mass-8 nuclei  
ensures there will be no astrophysically interesting production of  
any heavier nuclides.   
   
The primordial nuclear reactor is short-lived, quickly encountering an energy   
crisis.  Because of the falling temperature and the Coulomb barriers, nuclear   
reactions cease rather abruptly as the temperature drops below ~$\sim 30$~keV,   
when the Universe is $\sim 20$~minutes old.  This results in ``nuclear   
freeze-out,'' since no already existing nuclides are destroyed (except for 
those   that are unstable and decay) and no new nuclides are created.  In 
$\sim 1000$  seconds BBN has run its course.   
   
\subsection{The SBBN-predicted Abundances}   
   
The primordial abundances of D, \3he, and \7li($^7$Be) are rate limited,   
depending sensitively on the competition between the nuclear reaction   
rates (proportional to the nucleon density) and the universal expansion   
rate.  As a result, these nuclides are all potential baryometers.  As   
the Universe expands, the nucleon density decreases so it is useful to   
compare it to that of the CBR photons: $\eta \equiv n_{\rm N}/n_{\gamma}$.    
Since this ratio turns out to be very small, it is convenient to introduce   
\be   
\eta_{10} \equiv 10^{10}(n_{\rm N}/n_{\gamma}) = 274\Omega_{\rm b}h^{2}\,,   
\ee   
where \omb is the ratio of the present values of the baryon and critical 
densities and $h$ is the present value of the Hubble parameter in units of 
100~km s$^{-1}$ Mpc$^{-1}$ .  As the Universe evolves (post-\epm annihilation) 
this  ratio is accurately preserved so that $\eta$ at the time of BBN should  
be equal to its value today. Testing this relation over ten orders of  
magnitude in redshift, over a timespan of some 10 billion years, can  
provide a confirmation of, or pose a challenge to the standard model.    
   
\begin{figure*}[t]  
\includegraphics[width=1.0\columnwidth,angle=0,clip]{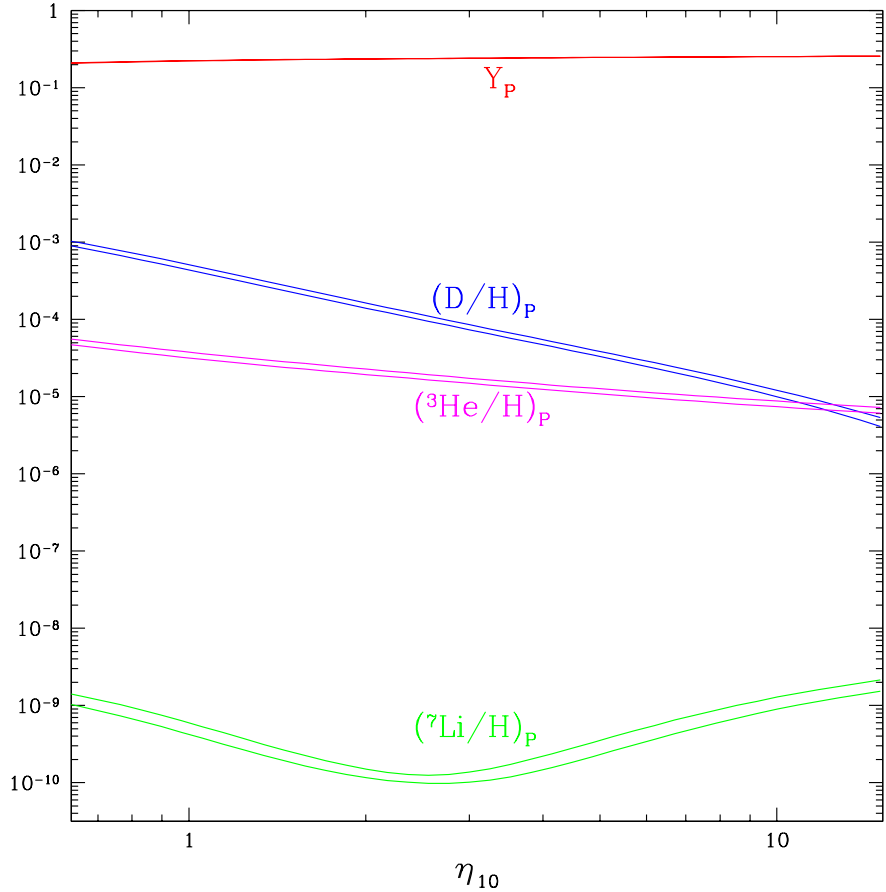}  
\vskip 0pt \caption{The SBBN-predicted primordial abundances of D,   
\3he, and \7li (by number with respect to hydrogen), and the \4he mass    
fraction Y as a function of the nucleon abundance $\eta_{10}$.    
The widths of the bands reflect the theoretical uncertainties.   
\label{fig:schrplot}}  
\end{figure*}

In contrast to the other light nuclides, the primordial abundance of   
\4he (mass fraction Y) is relatively insensitive to the baryon density,
but since virtually all neutrons available at BBN are incorporated in   
\4he, Y does depend on the competition between the weak interaction   
rates (largely fixed by the accurately measured neutron lifetime) and   
the universal expansion rate.  The higher the nucleon density, the   
earlier can the D bottleneck be breached.  Since at early times there   
are more neutrons (as a fraction of the nucleons), more \4he will be   
synthesized.  This latter effect is responsible for a very slow   
(logarithmic) increase in Y with $\eta$.  Given the standard model   
relation between time and temperature and the measured nuclear and   
weak cross sections and decay rates, the evolution of the light-nuclide   
abundances may be calculated and the relic, primordial abundances   
predicted as a function of the one free parameter,  the nucleon   
density or $\eta$.  These predictions for SBBN are shown in Figure 
\ref{fig:schrplot}.   
  
Not shown on Figure~\ref{fig:schrplot} are the relic abundances of $^6$Li,   
$^9$Be, $^{10}$B, and $^{11}$B; for the same range in $\eta$, all of   
them lie offscale, in the range $10^{-20} - 10^{-13}$.  The results   
shown here are from the BBN code developed and refined over the years   
by my colleagues at The Ohio State University (OSU).  They are in   
excellent agreement with the published results of the Chicago group 
(Burles, Nollett, \& Turner 2001).  Notice that the abundances appear 
in Figure~\ref{fig:schrplot} as bands.  These reflect the theoretical  
uncertainties in the predicted abundances.  For the OSU code the errors  
in D/H and \3he/H are at the $\sim 8\%$ level, while they are much  larger, 
$\sim 12\%$, for \7li.  Burles et al. (2001), in a reanalysis of the relevant  
published cross sections, have reduced the theoretical errors by roughly  
a factor of 3 for D and \3he and a factor of  2 for \7li. The  
reader may not notice the band shown for \4he, since the uncertainty  
in Y, dominated by the very small uncertainty in the neutron lifetime,  
is at only the $\sim 0.2\%$ level ($\sigma_{\rm Y} \approx 0.0005$).    
  
Based on the discussion above it is easy to understand the trends shown   
in Figure~\ref{fig:schrplot}.  D and \3he are burned to \4he.  The higher   
the nucleon density, the faster this occurs, leaving behind fewer nuclei   
of D or \3he.  The very slight increase of Y with $\eta$ is largely due   
to BBN starting earlier at higher nucleon density (more complete burning   
of D, $^3$H, and \3he to \4he) and higher neutron-to-proton ratio (more   
neutrons, more \4he). The behavior of \7li is more interesting.  At   
relatively low values of $\eta_{10} ~\la 3$, mass-7 is largely synthesized   
as \7li [by $^3$H($\alpha$,$\gamma$)\7li reactions], which is easily   
destroyed in collisions with protons.  So, as $\eta$ increases at low   
values, destruction is faster and \7li/H {\it decreases}.  In contrast, at  
relatively high values of $\eta_{10} ~\ga 3$, mass-7 is largely synthesized   
as $^7$Be [via \3he($\alpha$,$\gamma$)$^7$Be reactions], which is more   
tightly bound than \7li and, therefore, harder to destroy.  As $\eta$   
increases at high values, the abundance of $^7$Be {\it increases}.    
Later in the evolution of the Universe, when it is cooler and neutral   
atoms begin to form, $^7$Be will capture an electron and $\beta$-decay   
to \7li.   
  
\subsection{Nonstandard BBN}  
  
The predictions of the primordial abundance of \4he depend sensitively on  
the early expansion rate (the Hubble parameter $H$) and on the amount---if  
any---of a $\nu_{e} - \bar\nu_{e}$ asymmetry (the $\nu_{e}$ chemical potential  
$\mu_{e}$ or the neutrino degeneracy parameter $\xi_{e}$).  In contrast to  
\4he, the BBN-predicted abundances of D, \3he and \7li are determined by the  
competition between the various two-body production/destruction rates and the  
universal expansion rate.  As a result, the D, \3he, and \7li abundances are  
sensitive to the post-\epm annihilation expansion rate, while that of \4he  
depends on {\it both} the pre- and post-\epm annihilation expansion rates;  
the former determines the ``freeze-in'' and the latter modulates the importance  
of $\beta$-decay (see, \eg Kneller \& Steigman 2003).  Also, the primordial  
abundances of D, \3he, and \7li, while not entirely insensitive to neutrino  
degeneracy, are much less affected by a nonzero $\xi_{e}$ (\eg Kang \&  
Steigman 1992).  Each of these nonstandard cases will be considered below.   
Note that the abundances of at least two different relic nuclei are needed  
to break the degeneracy between the baryon density and a possible nonstandard  
expansion rate resulting from new physics or cosmology, and/or a neutrino asymmetry.  
     
\subsubsection{Additional Relativistic Energy Density}  
  
The most straightforward variation of SBBN is to consider the effect of  
a nonstandard expansion rate $H' \neq H$.  To quantify the deviation from 
the standard model it is convenient to introduce the ``{\it expansion rate 
factor}'' (or speedup/slowdown factor) $S$, where 
\be 
S \equiv H'/H = t/t'. 
\ee 
Such a nonstandard expansion rate might result from the presence of ``extra''  
energy contributed by new, light (relativistic at BBN) particles ``$X$''.  
These might, but need not, be additional flavors of active or sterile  
neutrinos.  For $X$ particles that are decoupled, in the sense that they  
do not share in the energy released by \epm annihilation, it is convenient  
to account for the extra contribution to the standard-model energy density  
by normalizing it to that of an ``equivalent'' neutrino flavor (Steigman et al.
1977),  
\be  
\rho_{X} \equiv \Delta N_{\nu}\rho_{\nu} =  
{7 \over 8}\Delta N_{\nu}\rho_{\gamma}.  
\label{eq:deln}  
\ee  
For SBBN, \Deln = 0 ($N_{\nu} \equiv 3 + \Delta N_{\nu}$) and for each   
such additional ``neutrino-like'' particle (\ie any two-component fermion),   
if $T_{X} = T_{\nu}$, then \Deln = 1; if $X$ should be a scalar, \Deln =   
4/7.  However, it may well be that the $X$ have decoupled even earlier in   
the evolution of the Universe and have failed to profit from the heating   
when various other particle-antiparticle pairs annihilated (or unstable   
particles decayed).  In this case, the contribution to \Deln from each   
such particle will be $< 1$ ($< 4/7$).  Henceforth we drop the $X$   
subscript.  Note that, in principle, we are considering any term in the   
energy density that scales like ``radiation'' (\ie decreases with the   
expansion of the Universe as the fourth power of the scale factor).  In   
this sense, the modification to the usual Friedman equation due to higher   
dimensional effects, as in the Randall-Sundrum model (Randall \& Sundrum  
1999a,b; see also Cline, Grojean, \& Servant 1999; Binetruy \etal 2000;
Bratt \etal 2002), may be included as well.  The interest in this latter 
case is that it permits the possibility of an apparent {\it negative} 
contribution to the radiation density ($\Delta N_{\nu} < 0$; $S < 1$).  
For such a modification to the energy density, the pre-\epm annihilation 
energy density in Equation \ref{eq:rho0} is changed to  
\be  
(\rho_{\rm R})_{pre} = {43 \over 8}\left(1 +  
{7\Delta N_{\nu} \over 43}\right)\rho_{\gamma}.  
\label{eq:rhoxpre}  
\ee  
  
Since any {\it extra} energy density ($\Delta N_{\nu} > 0$) speeds up 
the expansion of the Universe ($S > 1$), the right-hand side of the 
time-temperature relation in Equation \ref{eq:ttpre} is smaller by 
the square root of the factor in parentheses in Equation \ref{eq:rhoxpre}.  
\be  
S_{pre} \equiv (t/t')_{pre} = (1 + {7\Delta N_{\nu} \over 43})^{1/2}  
= (1 + 0.163\Delta N_{\nu})^{1/2}.  
\label{eq:sxpre}  
\ee 
 
In the post-\epm annihilation Universe the extra energy density is   
diluted by the heating of the photons, so that
\be  
(\rho_{\rm R})_{post} = 1.68\,(1 + 0.135\Delta N_{\nu})\rho_{\gamma}  
\label{eq:rhoxpos}  
\ee  
and  
\be  
S_{post} \equiv (t/t')_{post} = (1 + 0.135\Delta N_{\nu})^{1/2}.  
\label{eq:sxpos}  
\ee  
  
\begin{figure*}[t]  
\includegraphics[width=1.0\columnwidth,angle=0,clip]{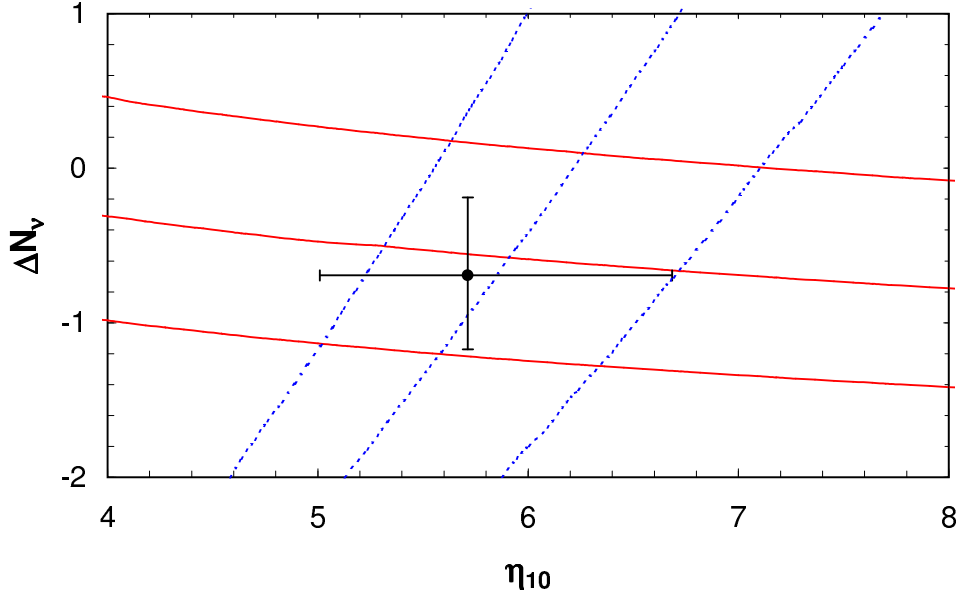}  
\vskip 0pt \caption{Isoabundance curves for D and \4he in the  
      \Deln -- $\eta_{10}$ plane.  The solid curves are  
      for  \4he (from top to bottom: Y = 0.25, 0.24, 0.23).  
      The dotted curves are for D (from left to right: \yd
      $ \equiv 10^{5}$(D/H) = 3.0, 2.5, 2.0).  The data point 
      with error bars corresponds to \yd = 2.6$\pm 0.4$ and 
      \Yp = 0.238$\pm 0.005$; see the text for discussion of 
      these abundances.  
\label{fig:nnuvseta}}  
\end{figure*}  
   
While the abundances of D, $^3$He, and $^7$Li are most sensitive to  
the baryon density ($\eta$), the $^4$He mass fraction (Y) provides  
the best probe of the expansion rate.  This is illustrated in Figure  
\ref{fig:nnuvseta} where, in the \Deln -- $\eta_{10}$ plane, are shown   
isoabundance contours for D/H and Y$_{\rm P}$ (the isoabundance curves  
for $^3$He/H and for $^7$Li/H, omitted for clarity, are similar in  
behavior to that of D/H).  The trends illustrated in Figure  
\ref{fig:nnuvseta} are easy to understand in the context of the   
discussion above.  The higher the baryon density ($\eta_{10}$),   
the faster primordial D is destroyed, so the relic abundance of   
D is {\it anticorrelated} with $\eta_{10}$.  But, the faster the   
Universe expands (\Deln $> 0$), the less time is available for   
D destruction, so D/H is positively, albeit weakly, correlated   
with $\Delta N_\nu$.  In contrast to D (and to \3he and $^7$Li),   
since the incorporation of all available neutrons into \4he is   
not limited by the nuclear reaction rates, the \4he mass fraction   
is relatively insensitive to the baryon density, but it is very   
sensitive to both the pre- and post-\epm annihilation expansion   
rates (which control the neutron-to-proton ratio).  The faster   
the Universe expands, the more neutrons are available for \4he.    
The very slow increase of \Yp with $\eta_{10}$ is a reflection   
of the fact that for a higher baryon density, BBN begins earlier,   
when there are more neutrons.  As a result of these complementary   
correlations, the pair of primordial abundances $y_{\rm D} \equiv   
10^{5}({\rm D/H})_{\rm P}$ and Y$_{\rm P}$, the \4he mass fraction, provide   
observational constraints on both the baryon density ($\eta$) and 
on the universal expansion rate factor $S$ (or on $\Delta N_{\nu}$)   
when the Universe was some 20 minutes old.  Comparing these to 
similar constraints from when the Universe was some 380 Kyr old,   
provided by the {\it WMAP} observations of the CBR polarization and   
the spectrum of temperature fluctuations, provides a test of   
the consistency of the standard models of cosmology and of  
particle physics and further constrains the allowed range of  
the present-Universe baryon density (\eg Barger \etal 2003a,b; 
Crotty, Lesgourgues, \& Pastor 2003; Hannestad 2003; Pierpaoli 2003).  
  
\subsubsection{Neutrino Degeneracy}

\begin{figure*}[t]  
\includegraphics[width=1.0\columnwidth,angle=0,clip]{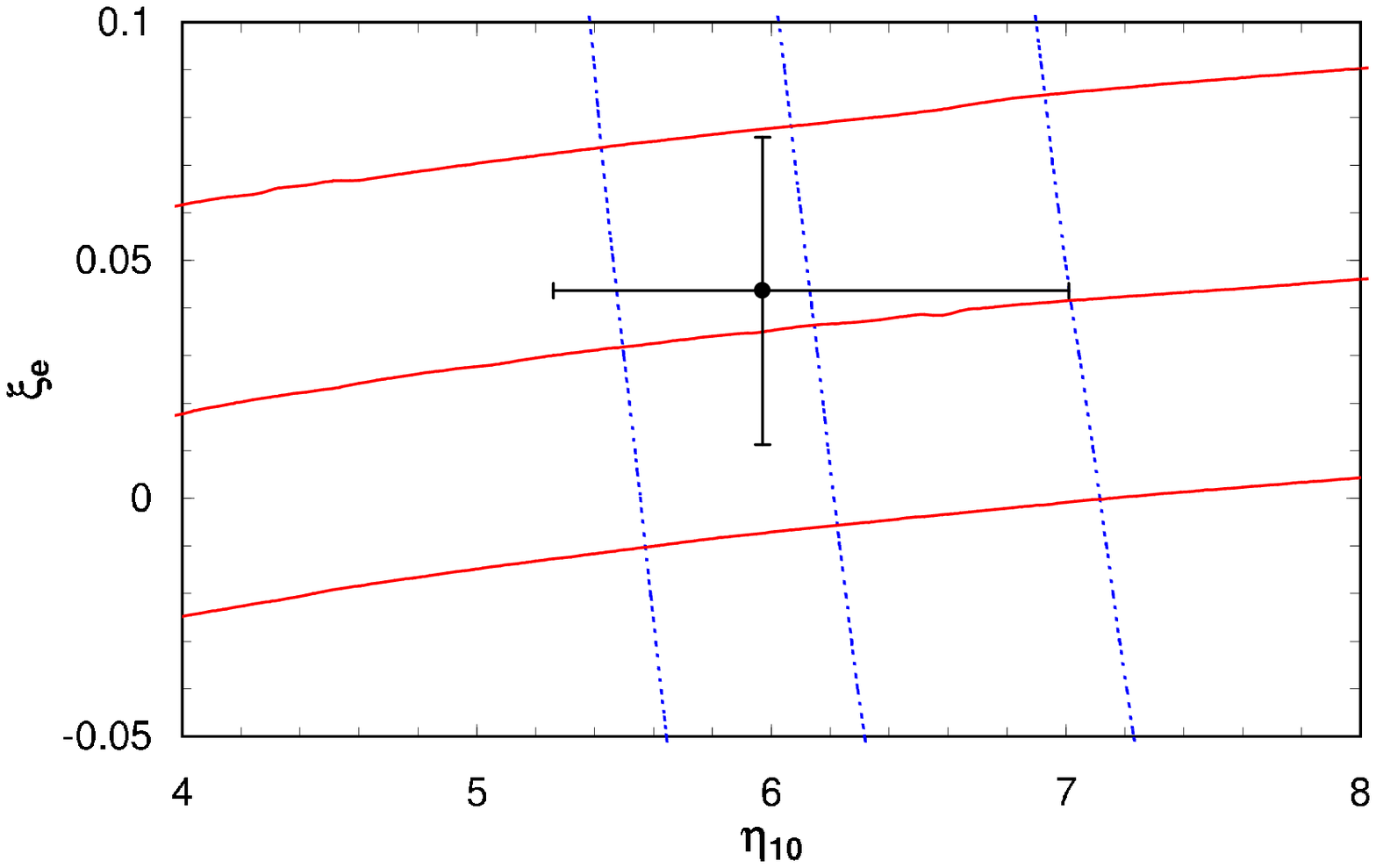}  
\vskip 0pt \caption{Isoabundance curves for D and \4he in the  
      $\xi_{e} - \eta_{10}$ plane.  The solid curves are  
      for  \4he (from top to bottom: \Yp = 0.23, 0.24, 0.25).  
      The dotted curves are for D (from left to right: \yd   
      $ \equiv 10^{5}$(D/H) = 3.0, 2.5, 2.0.)  The data point 
      with error bars corresponds to \yd = 2.6$\pm 0.4$ and 
      \Yp = 0.238$\pm 0.005$; see the text for discussion of 
      these abundances.  
   \label{fig:xivseta}}  
   \end{figure*}

The baryon-to-photon ratio provides a dimensionless measure of the  
universal baryon asymmetry, which is very small ($\eta ~\la 10^{-9}$).    
By charge neutrality the asymmetry in the charged leptons must also be   
of this order.  However, there are no observational constraints, save   
those to be discussed here (see Kang \& Steigman 1992; Kneller  
\etal 2002, and further references therein), on the magnitude of any  
asymmetry among the neutral leptons (neutrinos).  A relatively small  
asymmetry between electron type neutrinos and antineutrinos ($\xi_{e}  
~\ga 10^{-2}$) can have a significant impact on the early-Universe  
ratio of neutrons to protons, thereby affecting the yields of the  
light nuclides formed during BBN. The strongest effect is on the  
BBN \4he abundance, which is neutron limited.  For $\xi_{e} > 0$,  
there is an excess of neutrinos ($\nu_{e}$) over antineutrinos  
($\bar\nu_{e}$), and the two-body reactions regulating the neutron-to-proton 
ratio (Eq.~\ref{eq:betadecay}) drive down the neutron  
abundance; the reverse is true for $\xi_{e} < 0$.  The effect of a nonzero  
$\nu_{e}$ asymmetry on the relic abundances of the other light  
nuclides is much weaker.  This is illustrated in Figure~\ref{fig:xivseta}, 
which shows the D and \4he isoabundance curves in the $\xi_{e} -  
\eta_{10}$ plane.  The nearly horizontal \4he curves reflect  
the weak dependence of \Yp on the baryon density, along with its  
significant dependence on the neutrino asymmetry.  In contrast,  
the nearly vertical D curves reveal the strong dependence of \yd  
on the baryon density and its weak dependence on any neutrino  
asymmetry (\3he/H and \7li/H behave similarly: strongly dependent 
on $\eta$, weakly dependent on $\xi_{e}$).  This complementarity  
between \yd and \Yp permits the pair \{$\eta,\xi_{e}$\} to be  
determined once the primordial abundances of D and \4he are  
inferred from the appropriate observational data.  
  
\section{Primordial Abundances}  
  
It is clear from Figures~\ref{fig:schrplot} --~\ref{fig:xivseta} that tests  
of the consistency of SBBN, along with constraints on any new physics, will   
be data-driven.  While D (and/or \3he and/or \7li) largely constrain the   
baryon density and \4he plays a similar role for $\Delta N_\nu$ and/or   
for $\xi_{e}$, there is an interplay among $\eta_{10}$, $\Delta N_{\nu}$,  
and $\xi_{e}$, which is quite sensitive to the adopted abundances.  For  
example, a {\it lower} primordial D/H {\it increases} the BBN-inferred   
value of $\eta_{10}$, leading to a {\it higher} predicted primordial   
\4he mass fraction.  If the primordial \4he mass fraction derived from   
the data is ``low,'' then a low upper bound on \Deln (or a nonzero lower   
bound on $\xi_{e}$) will be inferred.  It is therefore crucial to avoid   
biasing any conclusions by {\it underestimating} the present uncertainties   
in the primordial abundances derived from the observational data.   
  
The four light nuclides of interest, D, \3he, \4he, and \7li follow   
very different evolutionary paths in the post-BBN Universe.  In addition,   
the observations leading to their abundance determinations are also very   
different.  Neutral D is observed in absorption in the UV; singly ionized    
\3he is observed in emission in Galactic \hii regions; both singly    
and doubly ionized \4he are observed in emission via recombinations   
in extragalactic \hii regions; \7li is observed in absorption in the   
atmospheres of very metal-poor halo stars. The different histories   
and observational strategies provide some insurance that systematic   
errors affecting the inferred primordial abundances of any one of the   
light nuclides are unlikely to distort the inferred abundances of the  
others.   
  
\subsection{Deuterium}  
  
The post-BBN evolution of D is straightforward.  As gas is incorporated   
into stars the very loosely bound deuteron is burned to \3he (and beyond).    
Any D that passes through a star is destroyed.  Furthermore, there are   
no astrophysical sites where D can be produced in an abundance anywhere   
near that observed (Epstein, Lattimer, \& Schramm 1976).  As a result,  
as the Universe evolves and gas is cycled through generations of stars,  
deuterium is only destroyed.  Therefore, observations of the deuterium  
abundance anywhere, anytime, provide {\it lower} bounds on its primordial  
abundance.  Furthermore, if D can be observed in ``young'' systems, in  
the sense of very little stellar processing, the observed abundance  
should be very close to the primordial value.  Thus, while there are  
extensive data on deuterium in the solar system and the local interstellar  
medium of the Galaxy, it is the handful of observations of deuterium  
absorption in high-redshift, low-metallicity  QSO absorption-line systems 
(QSOALS), which are potentially the most valuable.  At sufficiently 
high redshifts and low metallicities, the primordial abundance of 
deuterium should reveal itself as a ``deuterium plateau.''  
  
\begin{figure*}[t]  
\includegraphics[width=1.0\columnwidth,angle=0,clip]{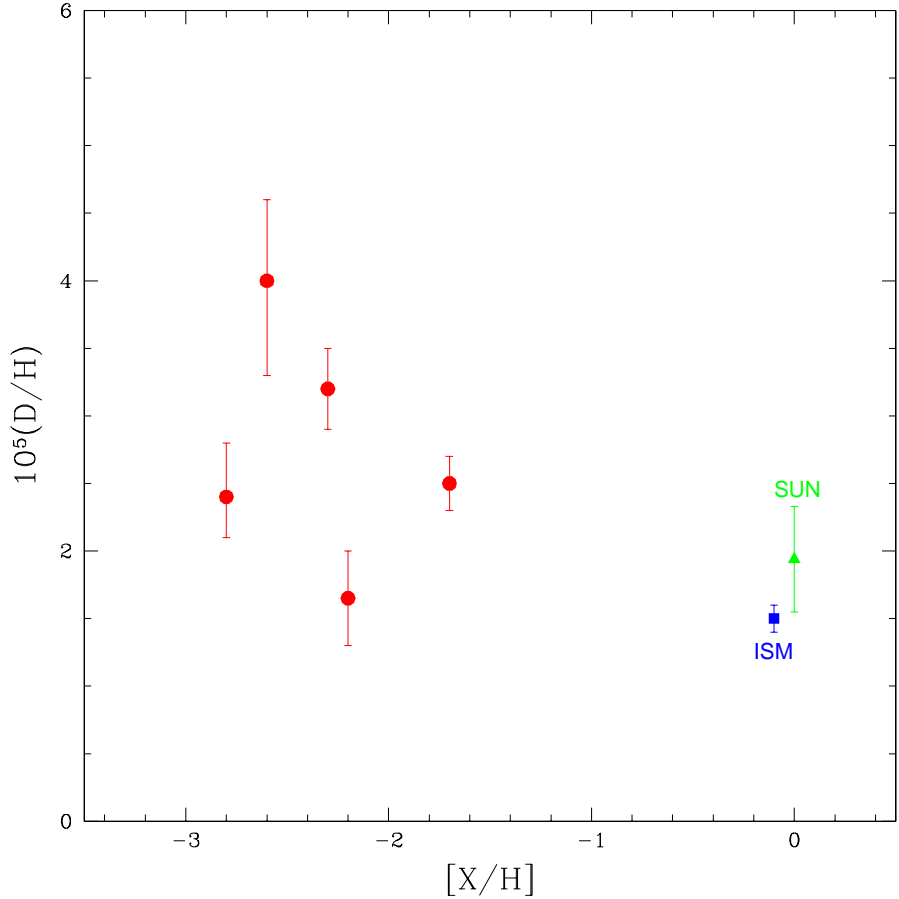}  
\vskip 0pt \caption{The deuterium abundance, D/H, versus metallicity,   
``X''(usually, X = Si), from observations (as of early 2003) of QSOALS 
(filled circles).  Also shown for comparison are the D abundances for 
the local ISM (filled square) and the solar system (``Sun''; filled 
triangle).    
\label{fig:dvssi}}  
\end{figure*}

\begin{figure*}[t]  
\includegraphics[width=1.0\columnwidth,angle=0,clip]{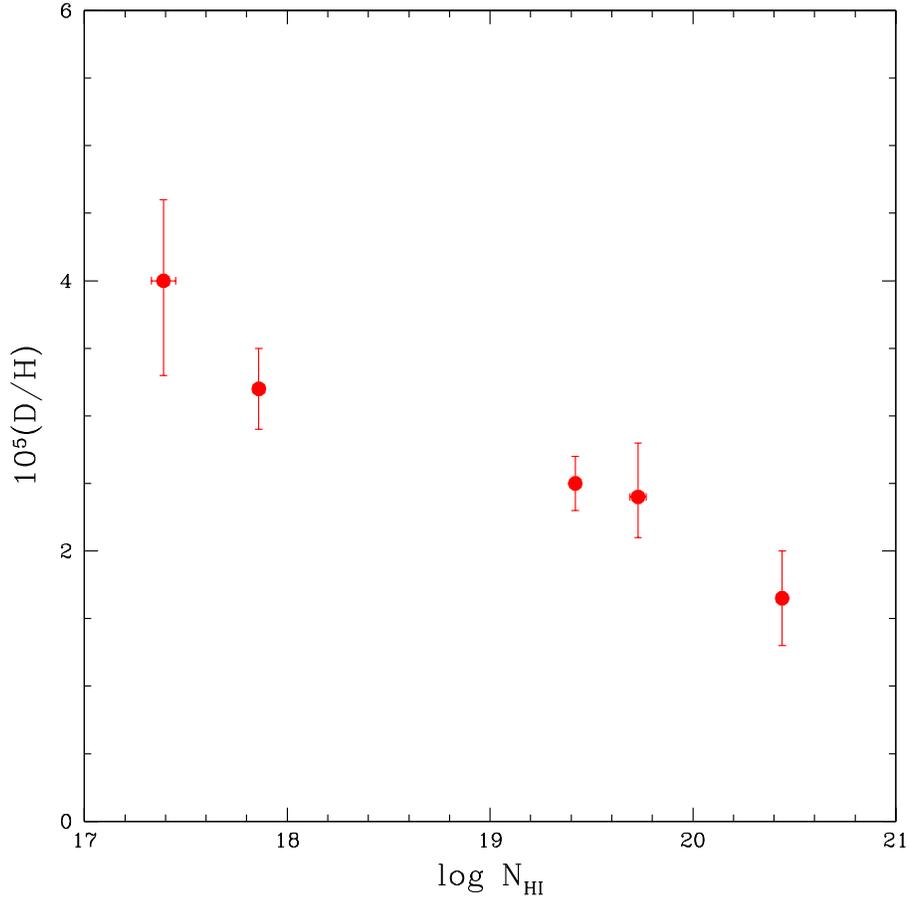}  
\vskip 0pt \caption{The deuterium abundance, D/H, versus the \hi 
column density in the absorbers, N(\hi), for the same QSOALS as  
in Figure~\ref{fig:dvssi}.    
\label{fig:dvsh}}  
\end{figure*}

\begin{figure*}[t]  
\includegraphics[width=1.0\columnwidth,angle=0,clip]{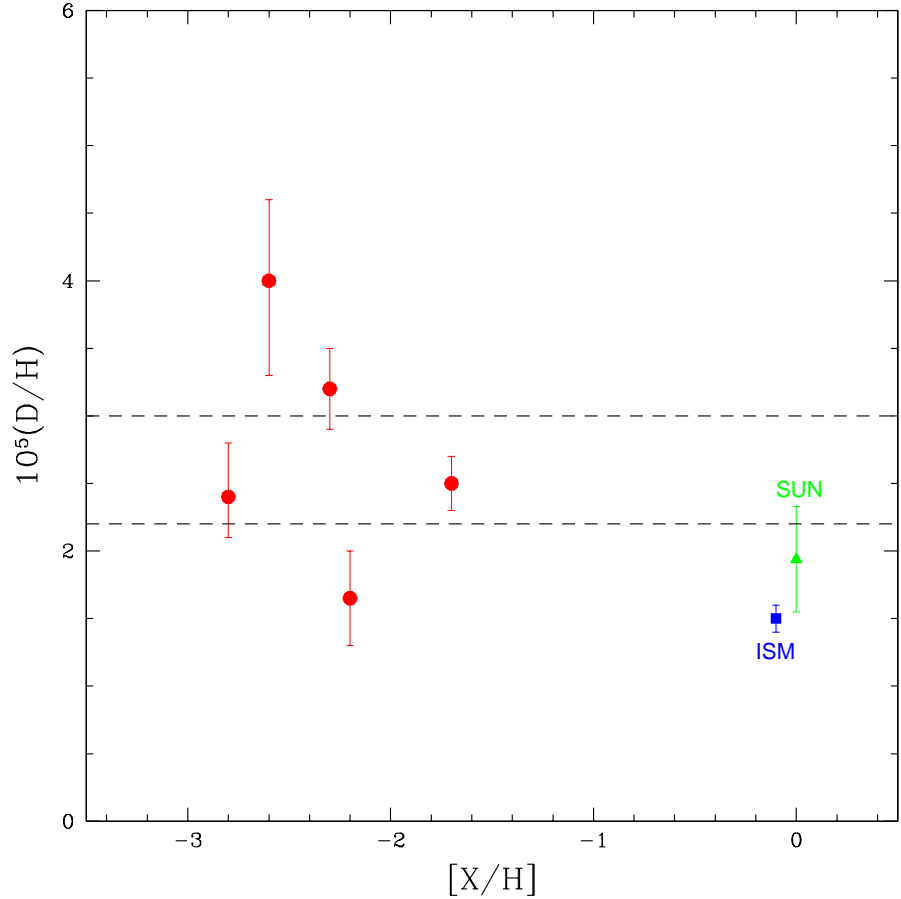}  
\vskip 0pt \caption{As in Figure~\ref{fig:dvssi}.  The dashed lines 
represent the $\pm 1 \sigma$ band calculated from the mean and its 
dispersion ((D/H)$_{\rm P} = 2.6 \pm 0.4 \times 10^{-5}$; see the text).    
\label{fig:dvssi03}}  
\end{figure*}

Inferring the primordial D abundance from the QSOALS has not been 
without its difficulties, with some abundance claims having been 
withdrawn or revised.  Presently there are ~$\sim$~half a dozen 
QSOALS with reasonably firm deuterium detections (Burles \& Tytler 
1998a,b; D'Odorico, Dessauges-Zavadsky, \& Molaro 2001; O'Meara \etal 
2001; Pettini \& Bowen 2002; Kirkman \etal 2003).  However, there is 
significant dispersion among the derived abundances, and the data fail 
to reveal the anticipated deuterium plateau (Fig.~\ref{fig:dvssi} -- 
\ref{fig:dvssi03}; see also Steigman 2003).  Furthermore, subsequent  
observations of the D'Odorico \etal (2001) QSOALS by Levshakov \etal 
(2002) revealed a more complex velocity structure and led to a 
revised---and uncertain---deuterium abundance.  This sensitivity to 
poorly constrained velocity structure in the absorbers is also exposed 
in the analyses of published QSOALS data by Levshakov and collaborators 
(Levshakov, Kegel, \& Takahara 1998a,b, 1999), which lead to consistent, 
but somewhat higher, deuterium abundances than those inferred from 
``standard'' data reduction analyses.   
  
Indeed, the absorption spectra of \di and \hi are identical, except for   
a wavelength/velocity offset resulting from the heavier reduced mass of   
the deuterium atom.  An \hi ``interloper,'' a low-column density cloud shifted 
by $\sim 81$~km s$^{-1}$ with respect to the main absorbing cloud, would   
masquerade as \di.  If this is not accounted for, a D/H ratio which is too   
high would be inferred.  Since there are more low-column density absorbers  
than those with high \hi column densities, absorption-line systems with   
somewhat lower \hi column density (\eg Lyman-limit systems) are more   
susceptible to this contamination than are the higher \hi column density  
absorbers (\eg damped Ly$\alpha$ absorbers).  However, for the damped 
Ly$\alpha$ absorbers,  
an accurate determination of the \hi column density requires an accurate  
placement of the continuum, which could be compromised by interlopers.  
This might lead to an overestimate of the \hi column density and a   
concomitant underestimate of D/H (J. Linsky, private communication).    
As will be seen, there is the possibility that each of these effects may  
have contaminated the current data.  Indeed, complex velocity structure  
in the D'Odorico \etal (2001) absorber (see Levshakov \etal 2002) renders  
it of less value in constraining primordial deuterium, and it will not  
be included in the estimates presented here.  
  
In Figure~\ref{fig:dvssi} are shown the extant data (circa June 2003) for  
D/H as a function of metallicity from the work of Burles \& Tytler  
(1998a,b), O'Meara \etal (2001), Pettini \& Bowen (2002), and Kirkman  
\etal (2003).  Also shown for comparison are the local interstellar medium 
(ISM) D/H (Linsky  \& Wood 2000) and that for the presolar nebula as inferred 
from solar  system data (Geiss \& Gloeckler 1998).   
  
On the basis of our discussion of the post-BBN evolution of D/H,   
a ``deuterium plateau'' at low metallicity was expected.  If, indeed,   
one is present, it is hidden by the dispersion in the current data.    
Given the possibility that interlopers may affect both the \di and 
the \hi column density determinations, it is interesting to plot 
D/H as a function of N(\hi).  This is shown in Figure~\ref{fig:dvsh}, 
where there is some (limited) evidence that D/H is higher in the  
Lyman-limit systems than in the damped Ly$\alpha$ absorbers.
  
To decide how to utilize this confusing data it may be of value to   
consider the observations chronologically.  Of the set chosen here,   
Burles \& Tytler (1998a,b) studied the first two lines of sight.   
For PKS 1937$-$1009 they derived \yd $\equiv 10^{5}$(D/H) = $3.25 \pm  
0.3$~(Burles \& Tytler 1998a), while for Q1009+299 they found \yd =   
$3.98^{+0.59}_{-0.67}$~(Burles \& Tytler 1998b).  These two determinations  
are in excellent agreement with each other ($\chi^{2} = 1.0$), leading  
to a mean abundance $\langle y_{\rm D}\rangle = 3.37 \pm 0.27$.  Next, 
O'Meara \etal (2001) added the line of sight to HS 0105+1619, finding a  
considerably lower abundance \yd = $2.54 \pm 0.23$.  Indeed, while  the 
weighted mean for these three lines of sight is $\langle y_{\rm D}\rangle =  
2.88$, the $\chi^{2}$ has ballooned to 6.4 (for two degrees of freedom).   
Absent any evidence that one or more of these abundances is in error,  
O'Meara \etal adopt the mean, and, for the error in the mean, they take  
the dispersion about the mean (0.72) divided by the square root of the number 
of data points: $\langle y_{\rm D}\rangle = 2.88 \pm 0.42$.  One year later  
Pettini \& Bowen (2002) published their {\it HST} data on the line of sight   
toward Q2206$-$199, finding a surprisingly low value of \yd = $1.65 \pm 0.35$. 
Including this determination reduces the mean to $\langle y_{\rm D}\rangle =  
2.63$, but the dispersion in \yd grows to 1.00 and $\chi^{2} = 16.3$ for  
three degrees of freedom.  Clearly, either one or more of these determinations  
is in error, or the variation among the high-redshift, low-metallicity   
deuterium abundances is larger than anticipated from our understanding   
of its evolution (Jedamzik \& Fuller 1997).  Using the mean and its  
dispersion (to fix the error), as of the time of the Carnegie Symposium, the 
best estimate for the primordial D abundance was $\langle y_{\rm D}\rangle =  
2.63 \pm 0.50$.  Shortly thereafter, in early 2003, the data of Kirkman  
\etal (2003) appeared for the line of sight toward Q1243+3047.  For  
this line of sight they find \yd = $2.42^{+0.35}_{-0.25}$.  This  
abundance lies between the lowest and the higher previous values,  
reducing the overall dispersion to 0.88, while hardly changing the mean  
from \yd = 2.63 to 2.60.  While the total $\chi^{2}$ is still enormous,  
increasing slightly to 16.6, the reduced $\chi^{2}$ decreases from 5.4  
to 4.2.  This is still far too large, suggesting that one or more of  
these determinations may be contaminated, or that there may actually be  
real variations in D/H at high redshifts and low metallicities.  Notice 
(see Fig.~\ref{fig:dvsh})  that the largest D/H estimates are from the two 
absorbers with the lowest  \hi column densities~(Lyman-limit systems),
where interlopers {\it might} contribute to  the inferred \di column 
densities, while the lowest abundances are from  the higher \hi column density 
(damped Ly$\alpha$ ) absorbers, where interlopers {\it  
might} affect the wings of the \hi lines used to fix the \hi column  
densities.  Absent any further data supporting, or refuting, these  
possibilities, there is no {\it a priori} reason to reject any of these  
determinations.  
 
To utilize the current data, the weighted mean D abundances for these   
five lines of sight and the dispersion are used to infer the abundance   
of primordial deuterium (and its uncertainty) adopted in this review:   
\yd = $2.6 \pm 0.4$.  Note that, given the large dispersion, two-decimal  
place accuracy seems to be wishful thinking at present.  For this reason,  
in quoting the primordial D abundance inferred from the observational  
data I have purposely chosen to quote values to only one decimal place.   
This choice is consistent too with the $\sim 3\%-8\%$ theoretical uncertainty  
(at fixed $\eta$) in the BBN-predicted abundance.  In Figure~\ref{fig:dvssi03}  
are shown the data, along with the corresponding $1 \sigma$ band.  It is  
worth remarking that using the same data Kirkman \etal (2003) derive a  
slightly higher mean D abundance: \yd = 2.74.  The reason for the difference  
is that they first find the mean of log(y$_{\rm D}$) and then use it to compute
the mean D abundance (\yd $ \equiv 10^{\langle \log(y_{\rm D})\rangle}$). 
  
\subsection{Helium-3}  
  
The post-BBN evolution of \3he is considerably more complex and model  
dependent than that of D.  Interstellar \3he incorporated into stars   
is burned to \4he (and beyond) in the hotter interiors, but preserved  
in the cooler, outer layers.  Furthermore, while hydrogen burning in   
cooler, low-mass stars is a net producer of \3he (Iben 1967; Rood 1972; 
Dearborn, Schramm, \& Steigman 1986; Vassiliadis \& Wood 1993; Dearborn,  
Steigman, \& Tosi 1996) it is unclear how much of this newly synthesized  
\3he is returned to the interstellar medium and how much of it is consumed  
in post-main sequence evolution (\eg Sackmann \& Boothroyd 1999a,b).   Indeed, 
it is clear that when the data (Geiss \& Gloeckler 1998; Rood \etal 1998; 
Bania, Rood, \& Balser 2002) are compared to a large variety of  
chemical evolution models~(Rood, Steigman, \& Tinsley 1976; Dearborn et al.
1996; Galli \etal 1997; Palla \etal 2000; Chiappini, 
Renda, \& Matteucci 2002), agreement is only possible for a very delicate  
balance between net production and net destruction of \3he.  For a recent  
review of the current status of \3he evolution, see Romano \etal (2003).   
Given this state of affairs it is not possible to utilize \3he as a  
baryometer, but it may perhaps be used to provide a consistency check.   
To this end, the abundance inferred by Bania \etal (2002) from an \hii  
region in the outer Galaxy, where post-BBN evolution might have been  
minimal, is adopted here: \y3 $ \equiv 10^{5}($\3he/H)$ ~= 1.1 \pm 0.2$.   
  
\subsection{Helium-4}   
   
Helium-4 is the second most abundant nuclide in the Universe   
after hydrogen.  In post-BBN evolution gas cycling though   
stars has its hydrogen burned to helium, increasing the \4he   
abundance above its primordial value.  As with deuterium, a   
\4he ``plateau'' is expected at sufficiently low metallicity.    
Although \4he is observed in the Sun and in Galactic \hii   
regions, the crucial data for inferring its primordial   
abundance is from observations of the helium and hydrogen    
emission (recombination) lines from low-metallicity, extragalactic  
\hii regions.  The present inventory of such regions studied  
for their helium content is approaching of order 100.  Thus,  
it is not surprising that even with modest observational  
errors for any individual \hii region, the statistical  
uncertainty in the inferred primordial abundance may be  
quite small.  In this situation, care must be taken with  
hitherto ignored or unaccounted for corrections and systematic  
errors or biases.   
   
\begin{figure*}[t]  
\includegraphics[width=1.0\columnwidth,angle=0,clip]{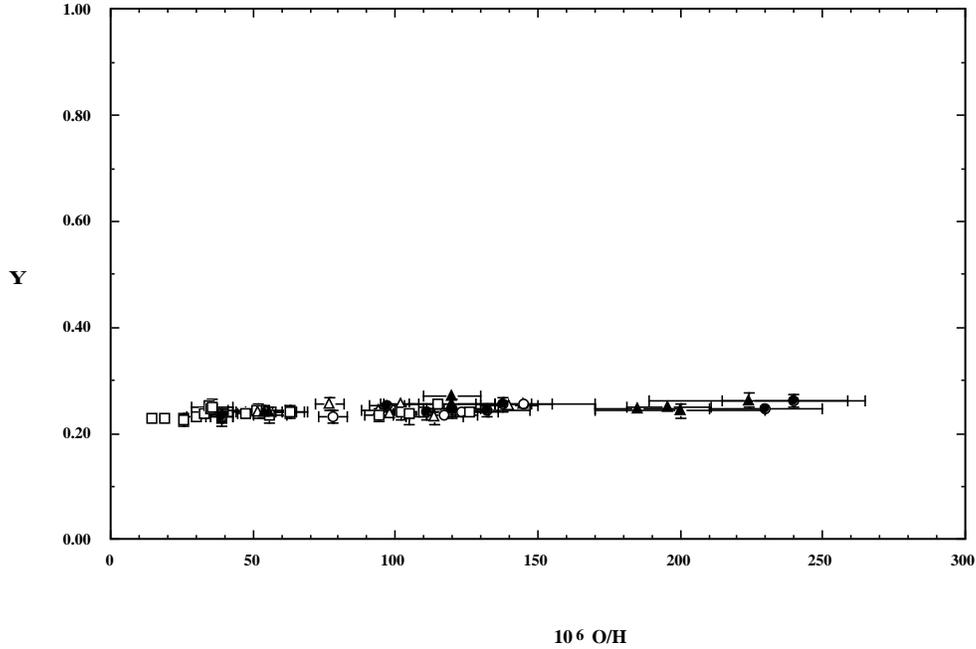}  
\vskip 0pt \caption{The \4he mass fraction, Y, inferred   
from observations of low-metallicity, extragalactic \hii   
regions versus the oxygen abundance derived from the same   
data. (Figure courtesy of K. A. Olive.)    
\label{fig:yvso}}   
\end{figure*}

In Figure~\ref{fig:yvso} is shown a compilation of the data used by   
Olive \& Steigman (1995) and Olive, Skillman, \& Steigman   
(1997), along with the independent data set obtained by  
Izotov, Thuan, \& Lipovetsky (1997) and Izotov \& Thuan  
(1998).  To track the evolution of the \4he mass fraction,   
Y is plotted versus the \hii region oxygen abundance.  These   
\hii regions are all metal poor, ranging from $\sim 1/2$ down   
to $\sim 1/40$ of solar (for a solar oxygen abundance of O/H   
$ \approx 5 \times 10^{-4}$; Allende-Prieto, Lambert, \& Asplund  
2001).  A key feature of Figure~\ref{fig:yvso} is that for sufficiently  
low metallicity the Y versus O/H relation approaches a \4he plateau!   
Since Y increases with metallicity, the relic abundance can either  
be bounded from above by the lowest metallicity regions, or the Y  
versus O/H relation may be extrapolated to zero metallicity.  The  
extrapolation is quite small, so that whether the former or the  
latter approach is adopted the difference in the inferred primordial  
abundance is small: $|\Delta$Y$|~ \la ~0.001$.   
   
While the data shown in Figure~\ref{fig:yvso} reveal a well-defined   
primordial abundance for \4he, the scale hides the very small  
statistical errors as well as the tension between the two groups'
helium abundances.  Olive \& Steigman (1995) and Olive et al. (1997) find 
Y$_{\rm P}$ $= 0.234 \pm 0.003$, but Izotov et al. (1997) and Izotov \& 
Thuan (1998) derive Y$_{\rm P}$ $=  0.244 \pm 0.002$.  
Although it is difficult to account for all  
of the difference, much of it is traceable to the different 
ways the two groups correct for the contribution to the  
emission lines from collisional excitation of neutral helium  
and also to Izotov and collaborators rejecting some helium emission lines 
{\it a  posteriori} when they yield ``too low'' an abundance.  Furthermore,  
for either data set, there are additional corrections for  
temperature, for temperature and density fluctuations, and for  
ionization, which when applied can change the inferred primordial  
\4he abundance by more than the quoted statistical errors (see,  
\eg Steigman, Viegas, \& Gruenwald 1997; Viegas, Gruenwald,
\& Steigman 2000; Gruenwald, Steigman, \& Viegas 2002; Peimbert, Peimbert 
\& Luridiana 2002; Sauer \& Jedamzik 2002). 
   
\begin{figure*}[t]  
\includegraphics[width=1.0\columnwidth,angle=0,clip]{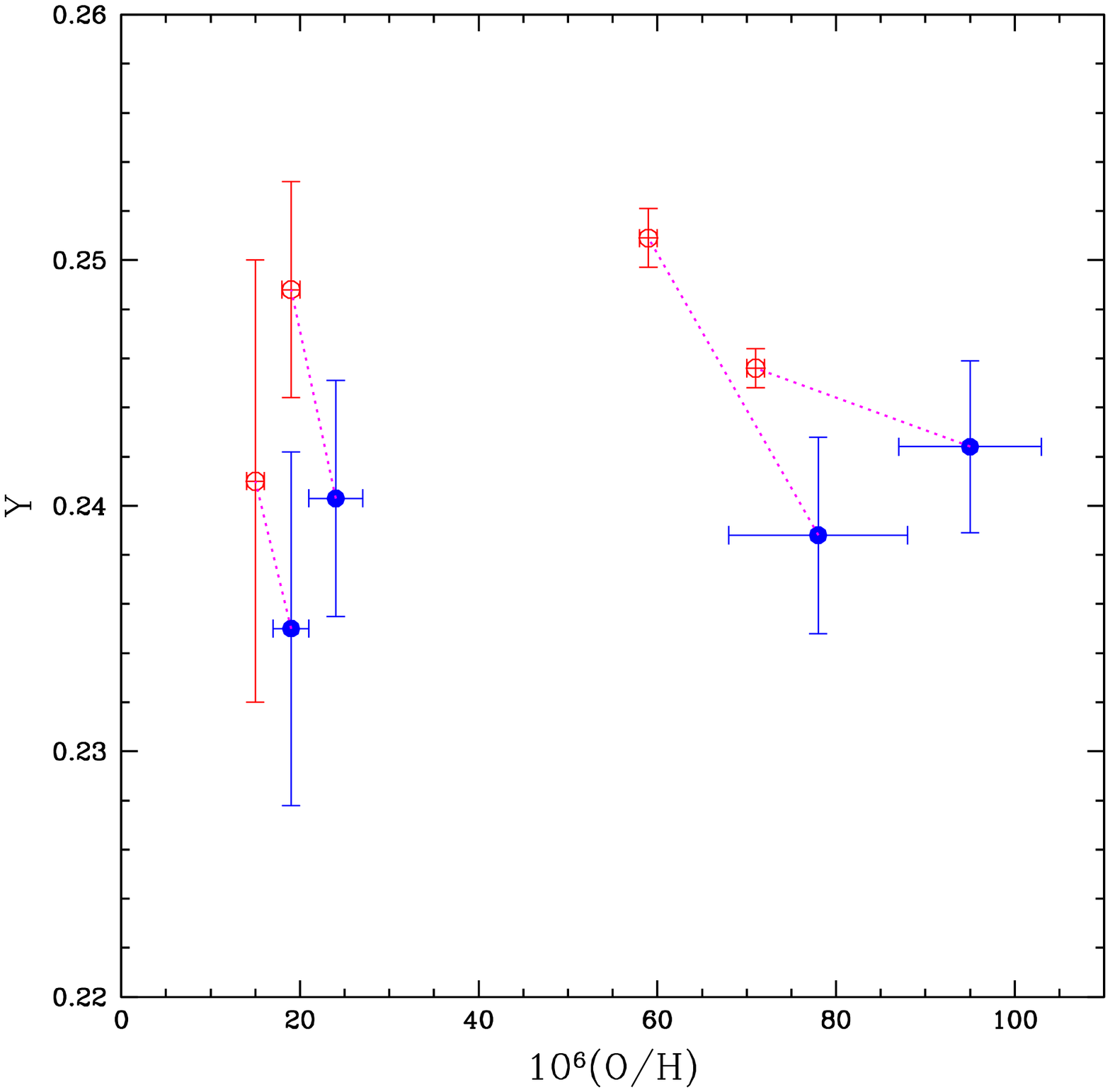}  
\vskip 0pt \caption{The Peimbert et al. (2002)   
reanalysis of the \4he abundance data for four of the Izotov \& 
Thuan (1998) \hii   regions.  The open circles are the Izotov \& Thuan (1998) 
abundances, while the filled circles are from Peimbert et al. (2002).
\label{fig:ppl}}  
\end{figure*}    
   
For example, Peimbert et al. (2002) recently reanalyzed the data from 
four of the Izotov \& Thuan (1998) \hii regions, employing their own 
\hii region temperatures and accounting for temperature fluctuations.  
Peimbert et al. (2002) derive systematically lower helium abundances, 
as shown in Figure~\ref{fig:ppl}.  From this very limited sample Peimbert 
et al. suggest that the Izotov \& Thuan (1998) estimate for the primordial 
\4he mass fraction might have to be reduced by as much as $\sim 0.007$.  
Peimbert et al. go further, combining their redetermined helium abundances 
for these four \hii regions with an accurate determination of Y in a 
more metal-rich \hii region (Peimbert, Peimbert, \& Ruiz 2000).   
Although these five data points are consistent with zero slope  
in the Y -- O/H relation, leading to a primordial abundance \Yp  
= 0.240$\pm 0.001$, this extremely small data set is also  
consistent with $\Delta$Y$ \approx 40($O/H), leading to a  
smaller primordial estimate of \Yp $\approx 0.237$.   
   
It seems clear that until new data address the unresolved   
systematic errors afflicting the derivation of the primordial   
helium abundance, the true errors must be much larger than the   
statistical uncertainties.  In an attempt to account for this,   
here I follow Olive, Steigman, \& Walker (2000) and adopt  
a compromise mean value along with a larger uncertainty:  
Y$_{\rm P} = 0.238 \pm 0.005$.   
 
\subsection{Lithium-7} 
 
\begin{figure*}[t]  
\includegraphics[width=0.85\columnwidth,angle=270,clip]{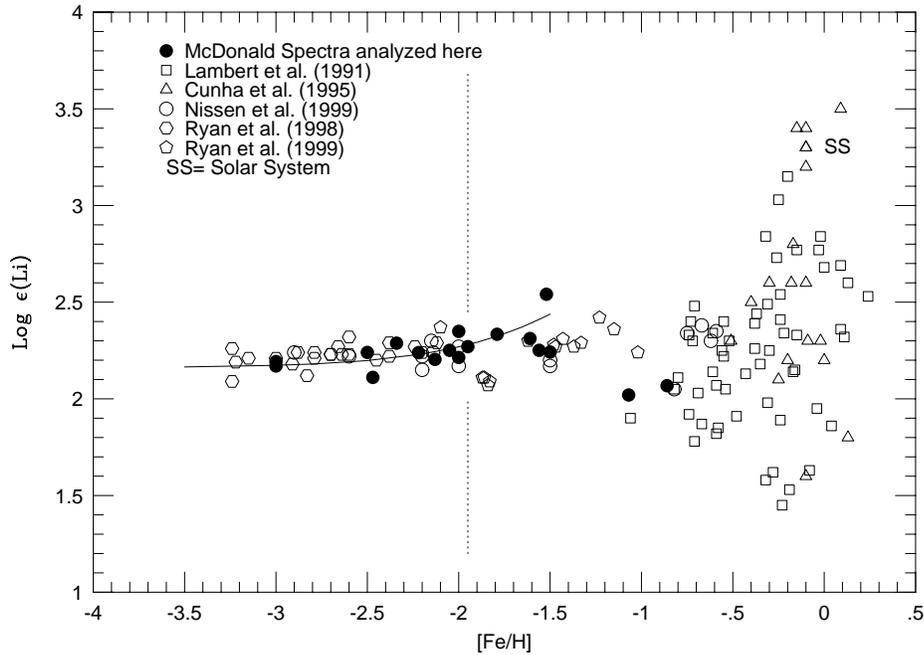}  
\vskip -30pt \caption{A compilation of the lithium abundance  
data as a function of metallicity from stellar observations  
(courtesy of V.~V.~Smith).  $\epsilon($Li$) \equiv 10^{12}$(Li/H),
and [Fe/H] is the usual logarithmic metallicity relative to 
solar.  Note the ``Spite plateau'' in Li/H for [Fe/H] $\la -2$. 
\label{fig:livsfe}}   
\end{figure*}

Lithium-7 is fragile, burning in stars at a relatively low 
temperature.  As a result, the majority of any interstellar  
\7li cycled through stars is destroyed.  For the same reason,  
it is difficult for stars to create new \7li and/or to return  
any newly synthesized \7li to the ISM before it is destroyed  
by nuclear burning.  In addition to synthesis in stars, the  
intermediate-mass nuclides $^6$Li, \7li, $^9$Be, $^{10}$B,  
and $^{11}$B can be synthesized via cosmic ray nucleosynthesis, 
either by alpha-alpha fusion reactions, or by spallation  
reactions (nuclear breakup) in collisions between protons and  
alpha particles and CNO nuclei.  In the early Galaxy, when  
the metallicity is low, the post-BBN production of lithium  
is expected to be subdominant to that from BBN abundance.   
As the data in Figure~\ref{fig:livsfe} reveal, only relatively  
late in the evolution of the Galaxy does the lithium abundance  
increase.  The data also confirm the anticipated ``Spite plateau''  
(Spite \& Spite 1982), the absence of a significant slope in  
the Li/H versus [Fe/H] relation at low metallicity due to the 
dominance of BBN-produced \7li.  The plateau is a clear signal  
of the primordial lithium abundance.  Notice, also, the enormous  
{\it spread} among the lithium abundances at higher metallicity.   
This range in Li/H likely results from the destruction/dilution  
of lithium on the surfaces of the observed stars while they are  
on the main sequence and/or lithium destruction during their  
pre-main sequence evolution, implying that it is the {\it upper  
envelope} of the Li/H versus [Fe/H] relation that preserves the  
history of Galactic lithium evolution.  Note, also, that at low  
metallicity the dispersion is much narrower, suggesting that  
corrections for depletion/dilution are (may be) much smaller  
for the Population II stars. 
 
As with the other relic nuclides, the dominant uncertainties  
in estimating the primordial abundance of \7li are not 
statistical, but systematic.  The lithium observed  
in the atmospheres of cool, metal-poor, Population II halo stars  
is most relevant for determining the BBN \7li abundance.   
Uncertainties in the lithium equivalent width measurements,  
in the temperature scales for the cool Population II stars, and in  
their model atmospheres dominate the overall error budget.   
For example, Ryan \etal (2000), using the Ryan, Norris, \&  
Beers (1999) data, infer [Li]$_{\rm P} \equiv ~12 +  
$log(Li/H)$~= 2.1$, while Bonifacio \& Molaro (1997) and  
Bonifacio, Molaro, \& Pasquini (1997) derive [Li]$_{\rm P} =  
2.2$, and Thorburn (1994) finds [Li]$_{\rm P} = 2.3$.  From  
recent observations of stars in a metal-poor globular cluster,  
Bonifacio \etal (2002) derive [Li]$_{\rm P} = 2.34 \pm 0.056$.   
As may be seen from Figure~\ref{fig:livsfe}, the indication  
from the preliminary data assembled by V. V. Smith (private  
communication) favors a Spite plateau at [Li]$_{\rm P}  
\approx 2.2$. 
 
In addition to these intrinsic uncertainties, there are 
others associated with stellar structure and evolution.  
The metal-poor halo stars that define the primordial  
lithium plateau are very old.  As a result, they have  
had time to disturb the prestellar lithium that could  
survive in their cooler, outer layers.  Mixing of these  
outer layers with the hotter interior where lithium has  
been (can be) destroyed will dilute or deplete the surface  
lithium abundance.  Pinsonneault \etal (1999, 2002) have 
shown that rotational mixing may decrease the surface  
abundance of lithium in these Population II stars by 0.1 -- 0.3  
dex while still maintaining the rather narrow {\it dispersion}  
among the plateau abundances (see also Chaboyer \etal 1992;  
Theado \& Vauclair 2001; Salaris \& Weiss 2002).  Pinsonneault  
\etal (2002) adopted for a baseline (Spite plateau) estimate  
[Li]$ ~= 2.2 \pm 0.1$, while for an overall depletion factor  
0.2 $\pm 0.1$ dex was chosen.  Adding these contributions to  
the log of the primordial lithium abundance {\it linearly},  
an estimate [Li]$_{\rm P} = 2.4 \pm 0.2$ was derived.  In  
the comparison between theory and observation below, I will  
adopt the Ryan \etal (2000) estimate [Li]$_{\rm P} = 2.1 \pm  
0.1$, but I will also consider the implications of the  
Pinsonneault \etal (2002) value. 

\section{Confrontation of Theory with Data} 
 
Having reviewed the basic physics and cosmological evolution 
underlying BBN and summarized the observational data leading 
to a set of adopted primordial abundances, the predictions may  
now be confronted with the data.  There are several possible  
approaches that might be adopted.  The following option is  
chosen here.  First, concentrating on the predictions of SBBN,  
deuterium will be used as the baryometer of choice to fix the 
baryon-to-photon ratio $\eta$.  This value and its uncertainty  
are then used to ``predict'' the \3he, \4he, and \7li abundances, 
which are compared to those adopted above.  This comparison can 
provide a test of the consistency of SBBN as well as identify 
those points of ``tension'' between theory and observation. 
This confrontation is carried further to consider the two 
extensions beyond the standard model [$S \neq 1$ (\Deln $\neq 
0$); $\xi_{e} \neq 0$]. 

\begin{figure*}[t]
\includegraphics[width=1.0\columnwidth,angle=0,clip]{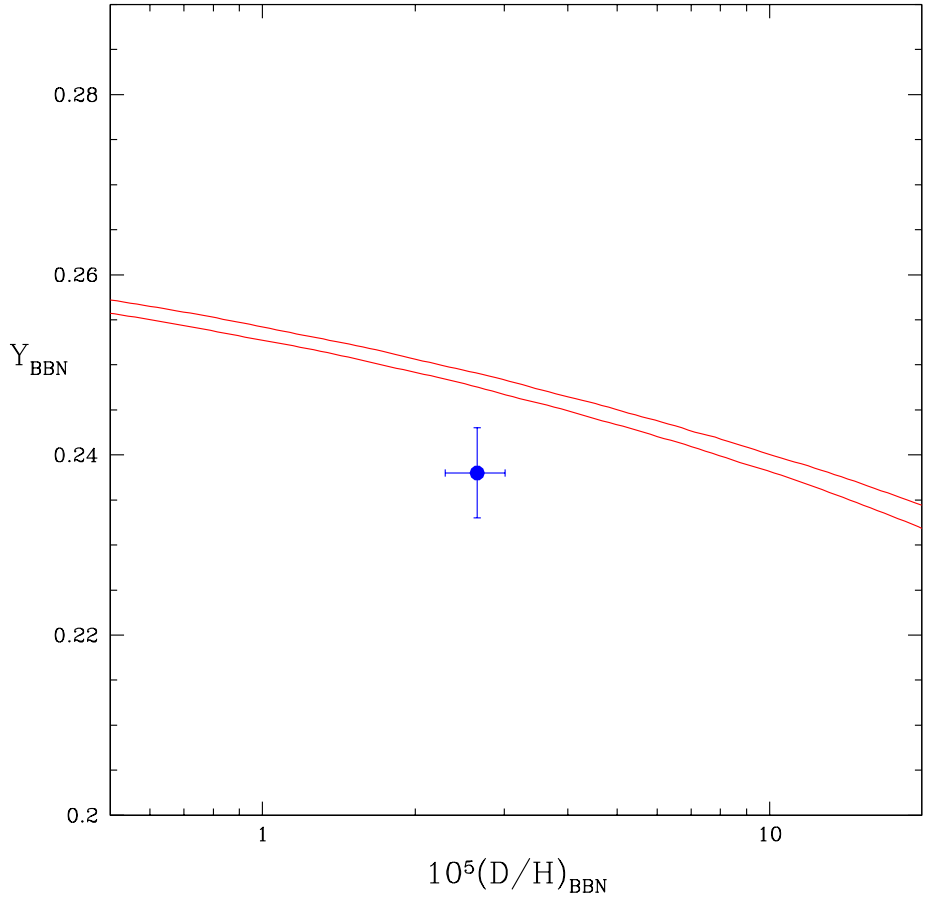}
\vskip 0pt \caption{The SBBN-predicted relation between the primordial
abundances of D and \4he (mass fraction) is shown by the band, whose
thickness represents the uncertainties in the predicted abundances.
Also shown by the point and error bars are the adopted primordial
abundances of D and \4he (see the text).
\label{fig:hevsd}}
\end{figure*}
 
\subsection{Testing the Standard Model} 
 
For SBBN, the baryon density corresponding to the D abundance  
adopted here ($y_{\rm D} = 2.6 \pm 0.4$) is $\eta_{10} = 6.1 
^{+0.7}_{-0.5}$, corresponding to \omb = $0.022^{+0.003}_{-0.002}$. 
This is in outstanding agreement with the estimate of Spergel  
\etal (2003), based largely on the new CBR ({\it WMAP}) data (Bennett  
\etal 2003): \omb $ = 0.0224 \pm 0.0009$.  For the baryon density 
determined by D, the SBBN-predicted abundance of \3he is $y_{3}  
= 1.0 \pm 0.1$, which is to be compared to the outer-Galaxy 
abundance of $y_{3} = 1.1 \pm 0.1$, which is suggested by Bania  
\etal (2002) to be nearly primordial.  Again, the agreement is  
excellent. 
 
\begin{figure*}[t]  
\includegraphics[width=1.0\columnwidth,angle=0,clip]{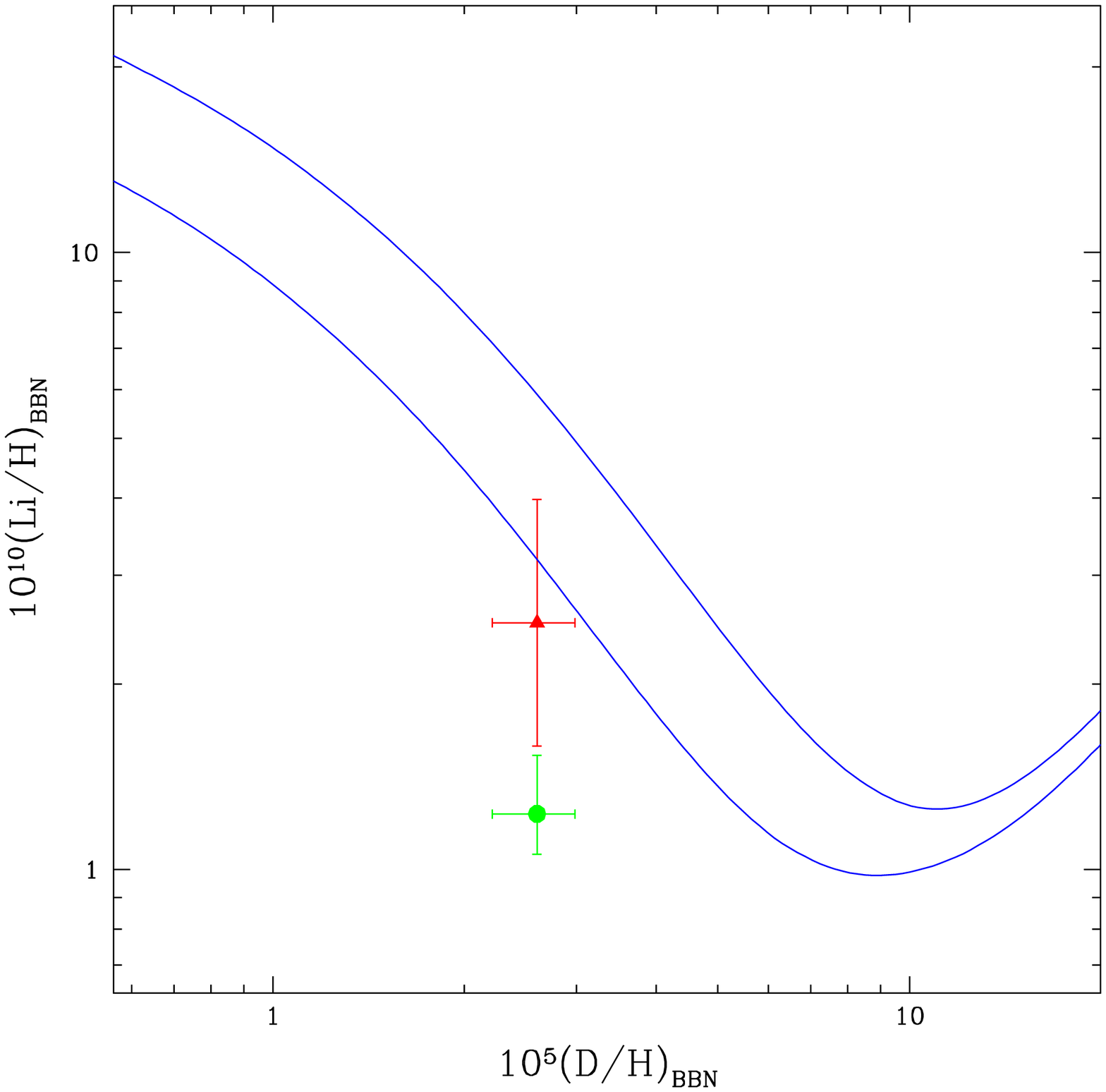}  
\vskip 0pt \caption{The SBBN-predicted relation between the primordial 
abundances of D and \7li is shown by the band, whose thickness  
reflects the uncertainties in the predicted abundances.  The 
data points are for the primordial abundance of D adopted here
along with the Ryan \etal (2000) Li abundance (filled circle) 
and the Pinsonneault \etal (2002) Li abundance (filled triangle).
\label{fig:livsd}}   
\end{figure*}   
 
The tension between the data and SBBN arises with \4he.  Given 
the very slow variation of \Yp with $\eta$, along with the very 
high accuracy of the SBBN-predicted abundance, the primordial 
abundance is tightly constrained: Y$_{\rm SBBN} = 0.248 \pm 0.001$. 
This should be compared with our adopted estimate of Y = $0.238 \pm 
0.005$ (Olive et al. 2000).  Agreement is only at the $\sim 5\%$ 
level. This tension is shown in Figure \ref{fig:hevsd}.  This apparent 
challenge to SBBN is also an opportunity.  As already noted, while  
the \4he abundance is insensitive to the baryon density, it is very 
sensitive to new physics (\ie nonstandard universal expansion  
rate and/or neutrino degeneracy). 
 
There is tension, too, when comparing the SBBN-predicted abundance of 
\7li with the (very uncertain) primordial abundance inferred from the 
data.  For SBBN the expected abundance is [Li]$_{\rm P} = 2.65^{+0.09}
_{-0.11}$.  This is to be compared with the various estimates above 
that suggested [Li]$_{\rm P} \approx 2.2 \pm 0.1$.  In Figure \ref
{fig:livsd} is shown the analog of Figure \ref{fig:hevsd} for lithium 
and deuterium.  Depending on the assessment of the uncertainty in the 
primordial abundance inferred from the observational data, the conflict 
with SBBN may or may not be serious.  In contrast to \4he, \7li is more 
similar to D (and to \3he) in that its BBN-predicted abundance is 
relatively insensitive to new physics.  As a result, this tension, if 
it persists, could be a signal of interesting new astrophysics (\eg 
have the halo stars depleted or diluted their surface lithium?). 
 
\subsection{Nonstandard Expansion Rate: $S \neq 1$ (\Deln $ \neq 0$)} 
 
\begin{figure*}[t]  
\includegraphics[width=1.0\columnwidth,angle=0,clip]{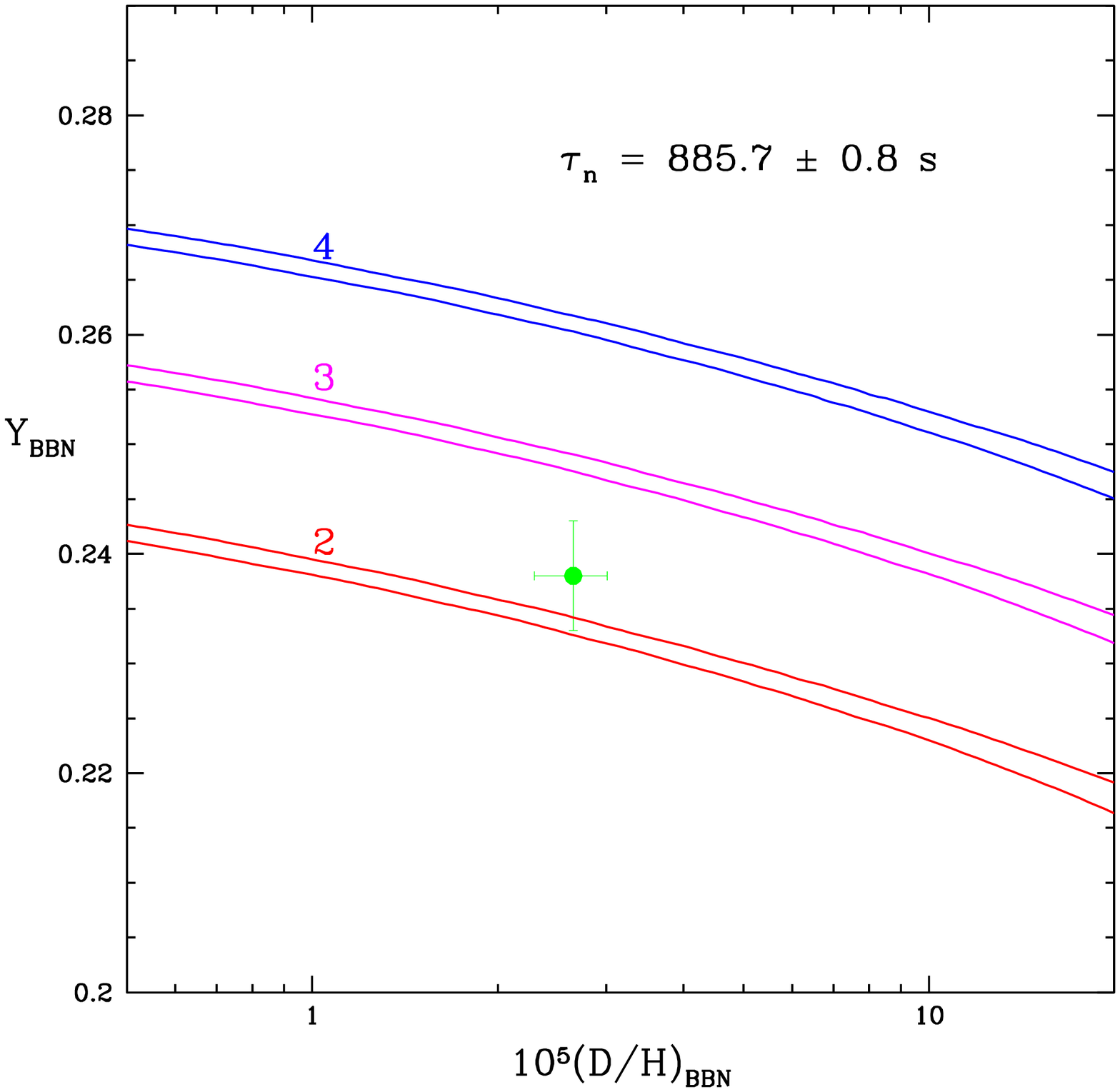}  
\vskip 0pt \caption{As in Figure \ref{fig:hevsd} for \Nnu = 2, 3, 4,  
which correspond to $S$ = 0.915, 1, 1.078. 
\label{fig:hevsd234}}   
\end{figure*}   
 
\begin{figure*}[t]  
\includegraphics[width=1.0\columnwidth,angle=0,clip]{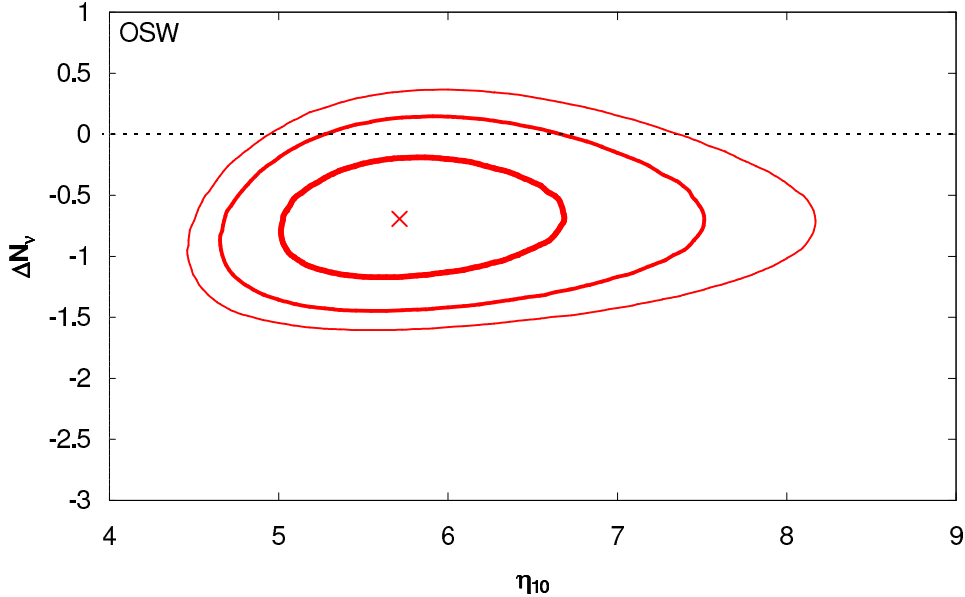}  
\vskip 0pt \caption{The 1-, 2-, and 3-$\sigma$ contours in the 
$\eta$ -- \Deln plane for BBN and the adopted D and \4he abundances. 
\label{fig:bbncontours}}   
\end{figure*}    

\begin{figure*}[t]  
\includegraphics[width=1.0\columnwidth,angle=0,clip]{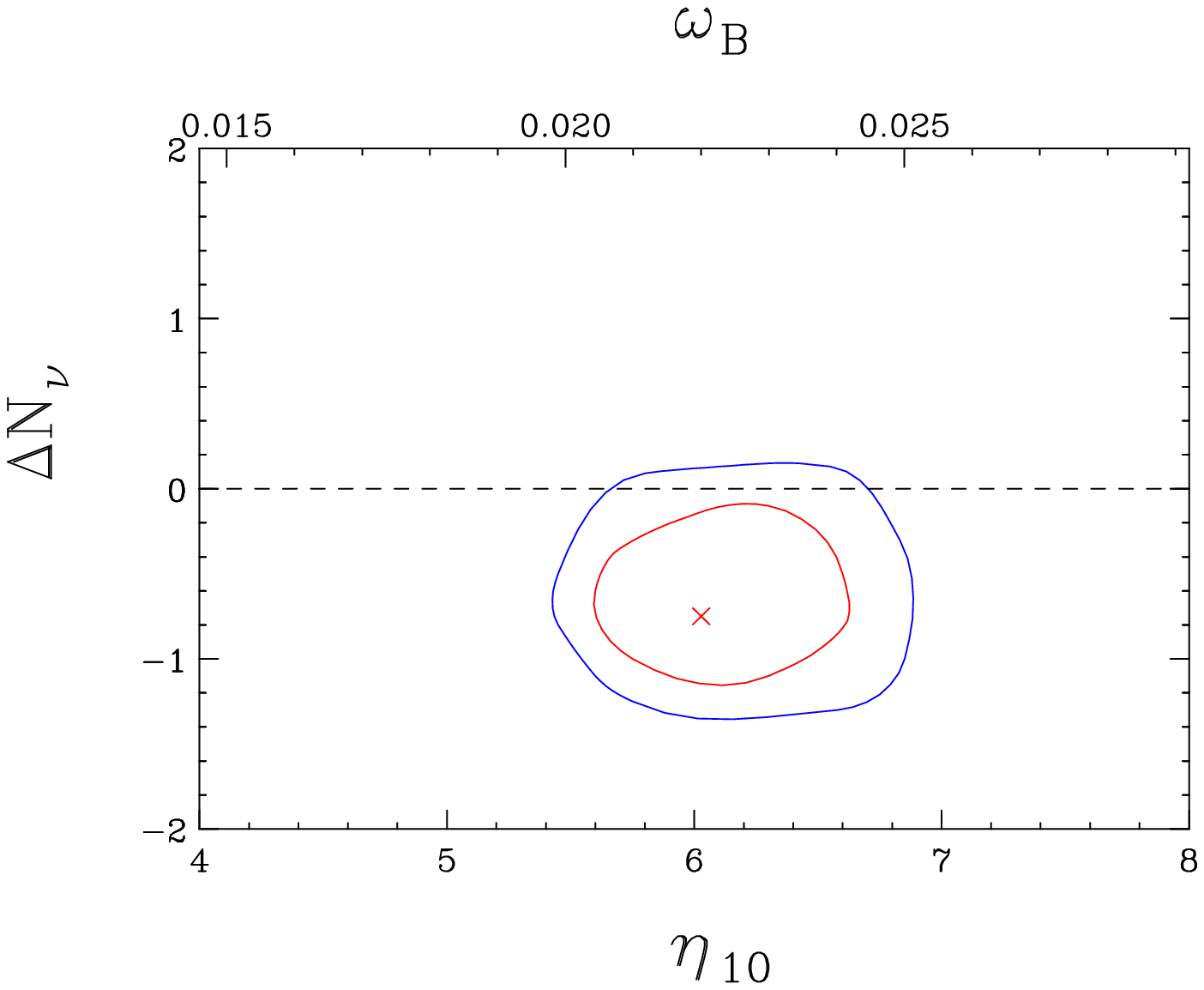}  
\vskip 0pt \caption{The 1- and 2-$\sigma$ contours in the $\eta$ --  
\Deln plane for the joint BBN -- CBR ({\it WMAP}) fit (Barger \etal 2003a). 
\label{fig:jointcontours}}   
\end{figure*}   
 
The excellent agreement between the SBBN-predicted baryon density 
inferred from the primordial-D abundance and that derived from 
the CBR and large scale structure (Spergel \etal 2003), and also 
the agreement between predicted and observed D and \3he suggest 
that the tension with \4he, if not observational or astrophysical 
in origin, may be a sign of new physics.  As noted earlier, \Yp 
is sensitive to the early-Universe expansion rate (while D, \3he, 
and \7li are less so).  A faster expansion ($S > 1$, \Deln > 0) 
leads to a higher predicted primordial abundance of \4he, and {\it vice versa }
for $S < 1$ (\Deln < 0).  In Figure \ref{fig:hevsd234} is shown the 
same \Yp versus \yd band as for SBBN in Figure \ref{fig:hevsd}, 
along with the corresponding bands for the nonstandard cases of 
a faster expansion (\Deln = 4) and a slower expansion (\Deln = 2).  
It can be seen that the data ``prefer'' a slower than standard early-Universe 
expansion rate.  If both $\eta$ and \Deln are allowed to  
be free, it is possible (not surprisingly) to accommodate the adopted  
primordial abundances of D and \4he (see Fig. \ref{fig:nnuvseta}).   
Given the similar effects of \Deln $\neq 0$ on the BBN-predicted D, 
\3he, and \7li abundances, while it is possible to maintain the good 
agreement (from SBBN) for \3he, the tension between \7li and D cannot 
be relieved.  In Figure \ref{fig:bbncontours} are shown the 1-, 2-,  
and 3-$\sigma$ BBN contours in the $\eta$ -- \Deln plane derived 
from the adopted values of \yd and Y$_{\rm P}$.  Although the best-fit point 
is at \Deln $= -0.7$ (and $\eta_{10} = 5.7$), it is clear 
that SBBN (\Nnu = 3) is acceptable. 
 
The CBR temperature anisotropy spectrum and polarization are also 
sensitive to the early-Universe expansion rate (see, \eg Barger 
\etal 2003a, and references therein).  There is excellent overlap  
between the $\eta$ -- \Deln confidence contours from BBN as shown in 
Figure \ref{fig:bbncontours} and from the CBR (Barger \etal 2003a).   
In Figure \ref{fig:jointcontours} are shown the confidence contours 
in the $\eta$ -- \Deln plane for a joint BBN -- CBR fit (Barger  
\etal 2003a).  Again, while the best fit value for \Deln is negative  
(driven largely by the adopted value for \Yp), \Deln = 0 is quite 
acceptable. 
 
\subsection{Neutrino Asymmetry ($\xi_{e} \neq 0$)} 
 
\begin{figure*}[t]  
\includegraphics[width=1.0\columnwidth,angle=0,clip]{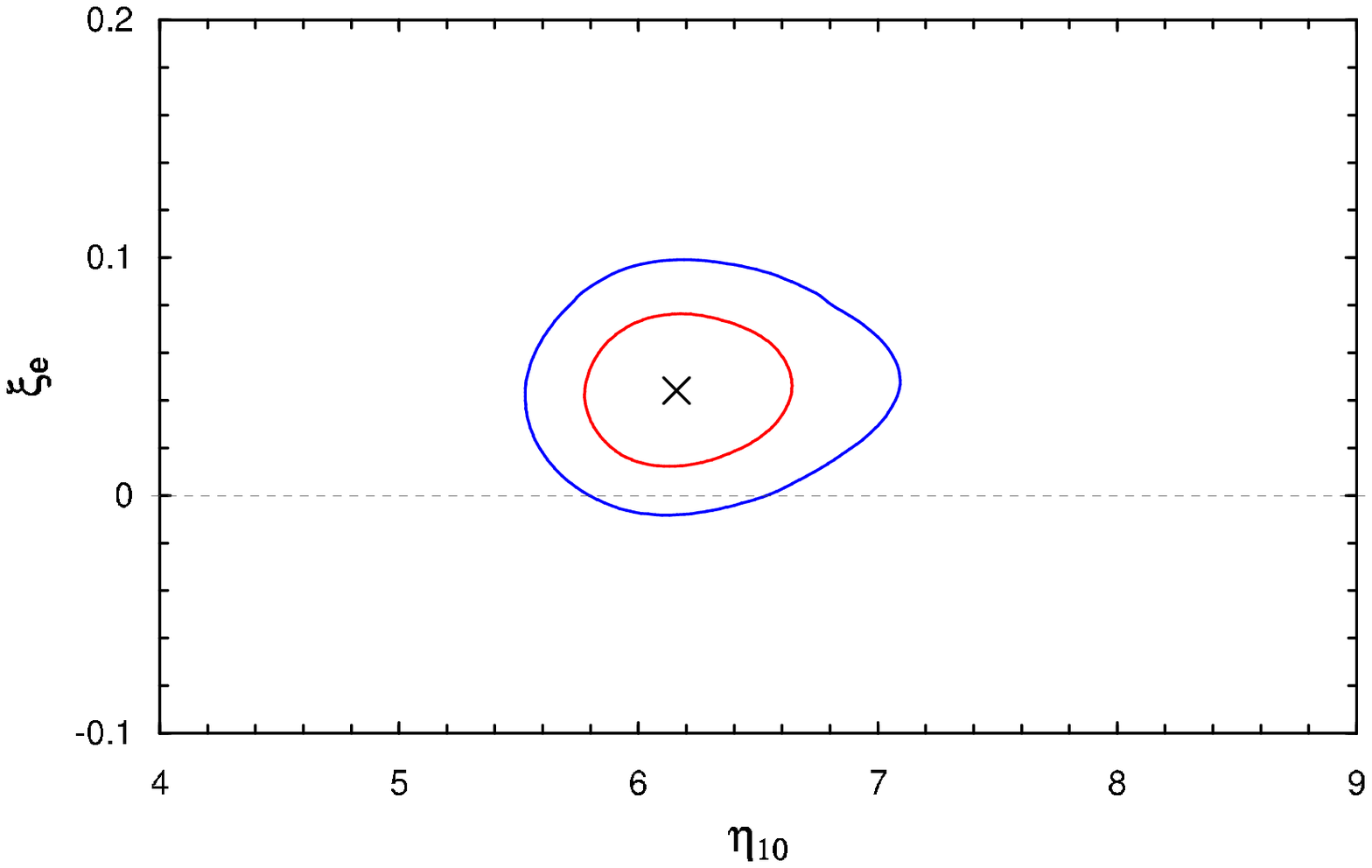}  
\vskip 0pt \caption{The 1-, 2-, and 3-$\sigma$ contours in the 
$\eta - \xi_{e}$ plane for BBN (\Nnu = 3) and the adopted D  
and \4he abundances (Barger \etal 2003b). 
\label{fig:xivsetacontours}}   
\end{figure*}   
 
The tension between D and \4he can also be relieved by nonstandard 
neutrino physics (see Fig. \ref{fig:xivseta}).  Although the asymmetry  
(difference between the numbers of particles and antiparticles) in  
charged leptons, tied to that in the baryons by charge neutrality  
of the Universe, must be very small, the neutrino asymmetry is  
unconstrained observationally.  Of relevance to BBN is the asymmetry  
between the electron neutrinos and the electron antineutrinos ($\xi_{e}$),
which regulates the pre-BBN neutron-to-proton ratio through the reactions  
in Equation~\ref{eq:betadecay}.  In Figure \ref{fig:xivsetacontours} are shown  
the 1- and 2-$\sigma$ contours in the $\eta - \xi_{e}$ plane for BBN (for  
\Nnu = 3) and the adopted abundances of D and \4he.  As seen before for  
\Deln $ \neq 0$, while a fit to the data can be achieved for $\xi_{e}  
\neq 0$, the data are not inconsistent with $\xi_{e} = 0$.  Furthermore, 
as is shown in Figure \ref{fig:xivsetacontours}, BBN constrains the  
allowed range for neutrino asymmetry to be very small.  For further  
implications for neutrino physics and for a discussion of the case  
where {\it both} \Deln and $\xi_{e}$ are free to differ from zero,  
see Barger \etal (2003b). 
 
\section{Summary and Conclusions} 
 
Given the standard models of cosmology and particle physics, SBBN 
predicts the primordial abundances of D, \3he, \4he, and \7li, which 
may be compared with the observational data.  Of the light nuclides, 
deuterium is the baryometer of choice, while \4he is an excellent 
chronometer.  The universal density of baryons inferred from SBBN 
and the adopted primordial D abundance is in excellent (exact!) agreement with 
that derived from non-BBN, mainly CBR data (Spergel \etal 2003): 
$\eta_{10}({\rm SBBN}) = 6.10^{+0.67}_{-0.52}$; $\eta_{10}({\rm CBR})  
= 6.14 \pm 0.25$.  For this baryon density, the predicted primordial 
abundance of \3he is also in excellent agreement with the (very uncertain) 
value inferred from observations of an outer-Galaxy \hii region (Bania  
\etal 2002).  In contrast, the SBBN-predicted mass fraction of \4he 
for the concordant baryon density is \Yp = $0.248 \pm 0.001$, while that 
inferred from observations of recombination lines in metal-poor, extragalactic 
\hii regions is lower (Olive et al. 2000): Y$_{\rm P}^{obs} = 0.238 
\pm 0.005$.  Since the uncertainties in the observationally inferred 
primordial value are likely dominated by systematics, this $\sim  
2\sigma$ difference may not be cause for (much) concern.  Finally, 
there appears to be a more serious issue concerning the predicted 
and observed lithium abundances.  While the predicted abundance is 
[Li]$_{\rm P} \approx 2.6 \pm 0.1$, current observations of metal-poor 
halo stars suggest a considerably smaller value $\approx 2.2 \pm 0.1$. 
 
It has been seen that the tension between D and \4he (or between the 
baryon density and \4he) can be relieved by either of two variations 
of the standard model (slower than standard early expansion rate; 
nonzero chemical potential for the electron neutrino).  However, 
in neither of these cases does the BBN-predicted \7li abundance move 
any closer to that inferred from the observations. 
 
In the current, data-rich era of cosmological research, BBN continues 
to play an important role.  The spectacular agreement in the baryon 
density inferred from processes occurring at widely separated epochs 
confirms the general features of the standard models of cosmology and 
particle physics.  The tensions with \4he and \7li provide challenges, 
and opportunities, to cosmology, to astrophysics, and to particle 
physics.  To outline these challenges and opportunities, let us 
consider each of the light nuclides in turn. 
 
For deuterium the agreement between SBBN and non-BBN determinations 
is perfect.  This may be surprising given the unexpectedly large 
dispersion among the handful of extant D abundance determinations at high 
redshifts and low metallicities.  Here, the challenge is to observers and 
theorists.  Clearly more data are called for.  Perhaps new data will reduce the 
dispersion.  In that case it can be anticipated that the SBBN-predicted 
baryon density will approach the accuracy of that currently available 
from non-BBN data.  On the other hand, newer data may support the 
dispersion, suggesting unexpectedly large variations in the D abundance 
at evolutionary times earlier than expected (Jedamzik \& Fuller 1997). 
Perhaps there is more to be learned about early chemical evolution. 
 
From studies of \3he in Galactic \hii regions (Balser \etal 1997;  
Bania \etal 2002) it appears that in the course of Galactic chemical  
evolution there has been a very delicate balance between post-BBN  
production and destruction.  If either had dominated, a gradient of  
the \3he abundance with galactocentric distance should have been seen  
in the data (see Romano \etal 2003, and references therein).  So far,  
none is.  Clearly, more data and a better understanding of the lower  
mass stars, which should dominate the production and destruction of 
\3he, would be of value. 
 
The very precise value of the baryon density inferred either from 
D and SBBN or from non-BBN data, coupled with the very weak dependence 
of the SBBN abundance of \4he on the baryon density, leads to a very  
precise prediction of its primordial mass fraction.  Although there 
exists a very large data set of \4he abundance determinations, the 
observational situation is confused at present.  It seems clear that 
while new data would be valuable, quality is much more important than 
quantity.  Data that can help resolve various corrections for temperature, 
for temperature and density fluctuations, for ionization corrections, 
would be of greater value than merely collecting more data that are 
incapable of addressing these issues.  Because of the very large data 
set(s), the {\it statistical} uncertainty in the derived primordial  
mass fraction is very small, $\sigma_{\rm Y_{\rm P}} \approx 0.002 -  
0.003$, while uncertain systematic corrections are much larger $\ga 
0.005$.  At this point it is systematics, not statistics that dominate  
the uncertainty in the primordial helium abundance.  In this context  
it is worth considering non-emission line observations that might 
provide an independent abundance determination.  Just such an 
alternative, the so-called R-parameter method using globular cluster 
stars was proposed long ago by Iben (1968) and by Iben \& Faulkner 
(1968).  It too has many systematic uncertainties associated with  
its application, but they are different from those for the emission-line 
studies.  Very recently, Cassisi, Salaris, \& Irwin (2003), 
using new stellar models and nuclear reactions rates, along with 
better data, find \Yp = $0.243 \pm 0.006$.  This is in much better 
agreement with the expected value (within $\la 1\sigma$) and should  
stimulate further investigations. 
 
The apparent conflict between the predicted and observed abundances  
of \7li, if not simply traceable to the statistical and systematic 
uncertainties, suggests a gap in our understanding of the structure 
and evolution of the very old, metal-poor, halo stars.  It would 
appear from the comparison between the predicted and observed 
abundances that lithium may have been depleted or diluted from 
the surfaces of these stars by $\sim 0.2 - 0.4$ dex.  Although  
a variety of mechanisms for depletion/dilution exist, the 
challenge is to account for such a large reduction without at  
the same time producing a large dispersion around the Spite plateau. 
 
The wealth of observational data accumulated over the last decade 
or more have propelled the study of cosmology from youth to 
maturity.  BBN has played, and continues to play, a central role in this 
process.  There have been many successes, but much remains to be 
done.  Whether the resolution of the current challenges are 
observational or theoretical, the future is bright. 
 
\vspace{0.3cm} 
{\bf Acknowledgements}. 
I am grateful to all the colleagues with whom I have worked, in the past 
as well as at present, for all I have learned from them; I thank 
them all.  Many of the quantitative results (and figures) presented 
here are from recent collaborations with V. Barger, J. P. Kneller, 
J. Linsky, D. Marfatia, K. A. Olive, R. J. Scherrer, S. M. Viegas, 
and T. P. Walker.  I thank V. V. Smith for permission to use Figure 
\ref{fig:livsfe}.  My research is supported at OSU by the DOE through 
grant DE-FG02-91ER40690.

\end{document}